\RequirePackage{tikz}
\documentclass[pdflatex,sn-mathphys]{sn-jnl}

\usepackage{soul} 
\usepackage{dcolumn}
\newcolumntype{d}[1]{D{.}{.}{#1}}

\usepackage{acronym}
\acrodef{CFD}[CFD]{Computational Fluid Dynamics}
\acrodef{DG}[DG]{Discontinuous Galerkin}
\acrodef{FV}[FV]{Finite Volume}
\acrodef{FD}[FD]{Finite Difference}
\acrodef{DGSEM}[DGSEM]{Discontinuous Galerkin Spectral Element Method}
\acrodef{DNS}[DNS]{Direct Numerical Simulation}
\acrodef{LES}[LES]{Large Eddy Simulation}
\acrodef{RANS}[RANS]{Reynolds-averaged Navier-Stokes}
\acrodef{MUSCL}[MUSCL]{Monotonic Upstream-centered Scheme for Conservation Laws}
\acrodef{LEVM}[LEVM]{Linear Eddy Viscosity Model}
\acrodef{DRSM}[DRSM]{Differential Reynolds Stress Model}
\acrodef{STG}[STG]{Synthetic Turbulence Generator}
\acrodef{LE}[LE]{leading edge}
\acrodef{TE}[TE]{trailing edge}
\acrodef{HSV}[HSV]{horse shoe vortex}
\acrodef{PV}[PV]{passage vortex}
\acrodef{TSV}[TSV]{trailing shed vortex}
\acrodef{LECV}[LECV]{leading edge corner vortex}
\acrodef{PS}[PS]{pressure side}
\acrodef{SS}[SS]{suction side}
\acrodef{POD}[POD]{Proper Orthogonal Decomposition}
\acrodef{PIV}[PIV]{Particle Image Velocimetry}
\acrodef{LPT}[LPT]{low-pressure turbine}
\acrodef{WENO}[WENO]{weighted essentially non-oscillatory}
\acrodef{KH}[KH]{Kelvin-Helmholtz}
\acrodef{LIC}[LIC]{line integral convolution}
\acrodef{MSER}[MSER]{marginal standard error rule}
\acrodef{LSB}[LSB]{laminar separation bubble}

\newcommand{\Ttref}{T_{\mathrm{t}1, \mathrm{ref}}}
\newcommand{\ptref}{p_{\mathrm{t}1, \mathrm{ref}}}
\newcommand{\pref}{p_{2, \mathrm{ref}}}
\newcommand{\xbycax}{x/C_{\mathrm{ax}}}
\newcommand{\re}{\mathrm{Re}}
\newcommand{\Ma}{\mathrm{Ma}}
\newcommand{\isth}{{2, \mathrm{th}}}

\newcommand{\nodeopacity}{0.8}
\newcommand{\figref}[1]{Fig.~\ref{#1}}
\newcommand{\MenterSST}{Menter SST $k$-$\omega$}
\newcommand{\SSGLRRw}{SSG/LRR-$\omega$}

\newcommand*\circled[1]{\textcircled{\footnotesize{#1}}}

\newcommand{\includegraphicswithtikz}[3][]{\begin{tikzpicture}[]
        \node[inner sep=0] at (0,0) {\includegraphics[#1]{#2}};
        #3
    \end{tikzpicture}}

\newcommand{\simplenode}[2]{\draw (#1) node [circle,fill=white,fill opacity=\nodeopacity,text opacity=1,align=center,draw=black,font=\footnotesize,inner sep=1pt]{#2};}
\newcommand{\simplelabel}[2]{\draw (#1) node [fill=white,fill opacity=\nodeopacity,text opacity=1,align=center,draw=black,font=\footnotesize,inner sep=1pt]{#2};}
\newcommand{\namednode}[3]{\node[circle,fill=white,fill opacity=\nodeopacity,text opacity=1,align=center,draw=black,font=\footnotesize,inner sep=1pt] (#3) at (#1) {#2};}
\newcommand{\namedlabel}[3]{\node [fill=white,fill opacity=\nodeopacity,text opacity=1,align=center,draw=black,font=\footnotesize,inner sep=1pt] (#3) at (#1) {#2};}
\newcommand{\simplearrow}[2]{\draw [->] (#1) -- (#2);}
\newcommand{\subfigurelabel}[2]{\draw (#1) node [text opacity=1,align=right,font=\footnotesize,inner sep=1pt,anchor=north east]{#2};}

\jyear{2021}%

\theoremstyle{thmstyleone}%
\theoremstyle{thmstyletwo}%

\theoremstyle{thmstylethree}%

\begin{document}

\title[LES of an LPT cascade with turbulent end wall boundary layers]{Large eddy simulation of a low-pressure turbine cascade with turbulent end wall boundary layers}


\author*[1]{\fnm{Christian} \sur{Morsbach}}\email{christian.morsbach@dlr.de}
\author[1]{\fnm{Michael} \sur{Bergmann}}
\author[1]{\fnm{Adem} \sur{Tosun}}
\author[2]{\fnm{Bjoern F.} \sur{Klose}}
\author[1]{\fnm{Edmund} \sur{Kügeler}}
\author[3]{\fnm{Matthias} \sur{Franke}}

\affil*[1]{\orgdiv{Institute of Propulsion Technology}, \orgname{German Aerospace Center (DLR)}, \orgaddress{\street{Linder Höhe}, \postcode{51147} \city{Cologne}, \country{Germany}}}
\affil[2]{\orgdiv{Institute of Test and Simulation for Gas Turbines}, \orgname{German Aerospace Center (DLR)}, \orgaddress{\street{Am Technologiezentrum 5}, \postcode{86159} \city{Augsburg}, \country{Germany}}}
\affil[3]{\orgname{MTU Aero Engines AG}, \orgaddress{\street{Dachauer Str. 665}, \postcode{80995} \city{Munich}, \country{Germany}}}


\abstract{
    We present results of implicit large eddy simulation (LES) and different Reynolds-averaged Navier-Stokes (RANS) models of the MTU 161 low pressure turbine at an exit Reynolds number of 90,000 and exit Mach number of 0.6.
    The LES results are based on a high order discontinuous Galerkin method and the RANS is computed using a classical finite-volume approach.
    The paper discusses the steps taken to create realistic inflow boundary conditions in terms of end wall boundary layer thickness and free stream turbulence intensity.
    This is achieved by tailoring the input distribution of total pressure and temperature, Reynolds stresses and turbulent length scale to a Fourier series based synthetic turbulence generator.
    With this procedure, excellent agreement with the experiment can be achieved in terms of blade loading at midspan and wake total pressure losses at midspan and over the channel height.
    Based on the validated setup, we focus on the discussion of secondary flow structures emerging due to the interaction of the incoming boundary layer and the turbine blade and compare the LES to two commonly used RANS models.
    Since we are able to create consistent setups for both LES and RANS, all discrepancies can be directly attributed to physical modelling problems.
    We show that the both a linear eddy viscosity model and a differential Reynolds stress model coupled with a state-of-the-art correlation-based transition model fail, in this case, to predict the separation induced transition process around midspan.
    Moreover, their prediction of secondary flow losses leaves room for improvement as shown by a detailed discussion turbulence kinetic energy and anisotropy fields.
}




\maketitle

\section{Introduction}
In the aviation industry's path towards sustainability, engines and their efficiency play a crucial role.
It is well known, that one way of increasing the efficiency of a turbofan engine is to increase its bypass ratio, i.e.~to increase the amount of mass flow which does not pass through the core engine with its compressor, combustion chamber and turbine.
Since the outer engine diameter is often limited by the aircraft design, this can be achieved by reducing the core engine size.
It leads, however, to relatively larger secondary flow regions, which e.g.~develop at intersections of blades and end walls or in rotor blade tip gaps and contribute significantly to the overall aerodynamic losses of the machine~\cite{Denton1993}.
Hence, understanding these losses and the capability of lower fidelity \ac{CFD} approaches such as \ac{RANS} to predict them is essential for further development in the field.

There exist several review papers discussing secondary flows in compressors~\cite{Gbadebo2005,Taylor2016} and turbines~\cite{Langston2001} in detail, mainly relying on experimental results or \ac{RANS} based \ac{CFD}.
To extend the existing knowledge numerically, methods are required which have little to no modelling uncertainty.
\ac{LES} or \ac{DNS} can be used to this end since they resolve the relevant turbulent length and time scales and, therefore, offer a detailed picture of the unsteady flow physics.
Due to the unfavourable scaling of the computational costs with the Reynolds number of roughly $\mathrm{Re}^{3.5}$ for \ac{DNS} and $\mathrm{Re}^{2.5}$ for wall-resolved \ac{LES}~\cite{Choi2012}, these methods have only become more widely used in the recent years with increasing availability of computational resources and advancement in high-order discretisation methods.
Many \ac{LES} studies of turbomachinery flows focus on the \ac{LPT} due to its low Reynolds number regime of $10^5$.
To further reduce the simulation costs, the assumption of a statistically 2D flow at the midspan of a turbine blade is often made by applying periodic boundary conditions in the spanwise direction, e.g.~\cite{Sandberg2015}.
These simulations have been used to improve the understanding of midspan flow physics and investigate the performance of \ac{RANS} approaches~\cite{Michelassi2015}.
In view of the above argument, the next logical step is to conduct 3D simulations, including the end wall boundary layers.

Cui et al.~\cite{Cui2017} conducted an \ac{LES} study on the influence of the inflow boundary layer state on the secondary flow system in the T106A \ac{LPT} with parallel end walls in incompressible conditions at a Reynolds number of 160,000.
They compare a laminar and a turbulent end wall boundary layer with laminar free stream as well as unsteady wakes with secondary flows generated by an upstream blade row and provide a detailed discussion of the secondary vortices with focus on loss generation mechanisms.
The same blade profile was studied using \ac{LES} by Pichler et al.~\cite{Pichler2019} at a lower Reynolds number of 120,000 and a Mach number of 0.59.
This paper also discusses the effect of different turbulent boundary layer shapes prescribed at the inflow but at freestream turbulence levels of roughly 5\%.
They found that while the boundary layer shape has an effect on the penetration depth of the secondary flows into the free stream, the midspan flow was largely unaffected by the difference.
Marconcini et al.~\cite{Marconcini2019} use the same \ac{LES} dataset to engage in a more detailed discussion on the secondary flow system and its representation by state-of-the-art industrial \ac{RANS}.
At midspan, their \ac{RANS} results show a significantly too narrow and too deep wake leading to an overall underestimation of the wake losses, which can be explained by an underprediction of the shear stress.
They state that, in general, \ac{RANS} predicts a similar secondary flow system with vortices showing a greater spanwise penetration compared with the \ac{LES}.
The peak losses associated with the vortex cores are larger for \ac{RANS} but in a pitch-averaged comparison, both methods are comparable.

The MTU T161, considered in this work, is representative of high-lift \ac{LPT} airfoils used in modern jet engines~\cite{Gier2007}.
In contrast to the extensively investigated T106 cascade, it features diverging end walls as found in real engines, such that the flow cannot be studied using a simple spanwise periodic setup.
Its geometry and boundary conditions are in the public domain and it has been the subject of both experimental~\cite{Martinstetter2010} and numerical~\cite{Mueller-Schindewolffs2017,Rasquin2021,Iyer2021} investigations.
The numerical studies have focused on operating points with a Mach number of 0.6 and Reynolds numbers of 90,000 and 200,000 based on isentropic exit conditions.
Müller-Schindewolffs et al.~\cite{Mueller-Schindewolffs2017} performed a direct numerical simulation of a section of the profile in which the effect of the diverging end walls was modelled using inviscid walls (termed quasi 3D, Q3D).
Recently, results including the end wall boundary layers and obtained with a second order \ac{FV} code were presented, but the analysis was focused on the flow physics in the mid-section~\cite{Afshar2022}.
Various computations of the full 3D configuration were conducted using high-order codes during the EU project TILDA~\cite{Rasquin2021,Iyer2021}.
However, due to the specification of laminar end wall boundary layers and no freestream turbulence at the inflow, no satisfactory agreement with the experiment could be obtained.
Recently, Rosenzweig et al.~\cite{Rosenzweig2022} presented a numerical study of the T161 obtained with a high-order \ac{FD} solver.
It features freestream turbulence and a Blasius boundary layer profile at the inflow plane with sufficient development length to transition to turbulence before reaching the turbine blade.

With a full 3D \ac{LES} including appropriate turbulent end wall boundary layers and freestream turbulence obtained with a high order \ac{DG} method, we aim to provide a high-quality reference dataset for this configuration.
Because \ac{RANS} is still the primary \ac{CFD} tool in the design process of turbomachinery components, we compare our \ac{LES} results to two commonly used \ac{RANS} setups and focus on the accuracy of the representation of secondary flow features.
The paper is structured as follows:
we will first provide a brief summary of the high-order \ac{DG} method used to perform the \ac{LES}.
This will be followed by a detailed discussion of how to set up both \ac{LES} and \ac{RANS} simulations consistently and in accordance with experimental data in terms of operating point, freestream turbulence and end wall boundary layer development.
Finally, after demonstrating very good agreement between \ac{LES} and experiment, we will discuss the performance of both \ac{LEVM} and more advanced \ac{DRSM} \ac{RANS} approaches in predicting the midspan and end wall flow features.

\section{Numerical method}

DLR's flow solver for turbomachinery applications is employed for the presented simulations of MTU's T161 cascade.
The LES is performed using the high-order \ac{DG} solver of TRACE~\cite{Bergmann2018a, Bergmann2020}.
Here, the implicitly filtered compressible Navier-Stokes equations are, first, mapped into the reference system, ensuring free-stream preservation~\cite{Kopriva2016}, and are spatially discretised with the 6th-order \ac{DGSEM} on collocated Legendre-Gauss-Lobatto nodes.
The collocation of the integration and solution nodes gives rise to a highly efficient discretisation scheme, since many numerical operations can be omitted by design.
Adjacent elements are coupled via Roe's approximate Riemann solver~\cite{Roe1981} for the advective part.
The required stabilisation for turbulent flows is achieved by using a split formulation of the scheme following Gassner et al.~\cite{Gassner2016}, applying the kinetic-energy-preserving two-point volume flux of Kennedy and Gruber~\cite{Kennedy2008}.
The viscous terms are discretised using the Bassi-Rebay 1~\cite{Bassi1997} scheme with a central numerical flux at the element interfaces, which, in combination with a stable advective discretisation, was found to be stable without requiring an additional penalty term~\cite{Gassner2018}.
Building on the favourable dispersion and dissipation characteristics of the high-order \ac{DG} scheme~\cite{Moura2017, Moura2017a}, the effect of the unresolved scales in the \ac{LES} is modelled implicitly by the numerical scheme itself and without applying an explicit subgrid-scale model.
To advance in time, the strong-stability preserving third-order explicit Runge-Kutta scheme of Shu and Osher~\cite{Shu1988} is used.
Resolved turbulent scales are injected at the inflow boundary using a \ac{STG} method based on randomised Fourier modes~\cite{Shur2014}.
The fluctuating velocity is introduced using a Riemann boundary condition~\cite{Leyh2020} and, in contrast to the inner faces, no numerical flux function is used at the inflow faces.
At the outflow plane, one-dimensional non-reflecting boundary conditions following Schlüß et al.~\cite{Schluess2016} are applied.

For the \ac{RANS} simulations, we use TRACE's density-based, compressible \ac{FV} solver~\cite{Morsbach2016,Geiser2019}, which employs a \ac{MUSCL} scheme~\cite{Leer1979} with a Van Albada 1 limiter~\cite{Albada1982} in combination with Roe's approximate Riemann solver~\cite{Roe1981} to discretise the convective fluxes and central derivatives for the viscous fluxes to obtain second order accuracy in space.
Steady state solutions are obtained with an implicit dual time-stepping approach which solves the five conservation equations in a fully coupled manner.
The additional turbulence model equations are solved in a conservative but segregated manner~\cite{Morsbach2012} where the implicit matrix is constructed only per equation.
Menter's SST $k$-$\omega$ model is implemented in its version from 2003~\cite{Menter2003} but with the production term modified according to Kato and Launder~\cite{Kato1993}.
Laminar-to-turbulent transition is accounted for using the correlation-based two-equation $\gamma$-$\re_\mathrm{\theta}$ model~\cite{Langtry2009}.
The SSG/LRR-$\omega$ \ac{DRSM} originally devised by Eisfeld is used in its latest version~\cite{Cecora2014,Morsbach2016}.
It is coupled with the same transition model with the intermittency $\gamma$ not only used to modify the production and dissipation terms but also the pressure-strain term~\cite{Nie2018}.
The remaining parts of the turbulence and transition model are left unchanged.
In order to achieve a setup consistent with \ac{LES}, the same formulation of inflow and outflow boundary conditions is used for \ac{RANS}.

\section{Setup and boundary conditions}
\subsection{Computational domain and mesh}
\begin{figure}
    \centering
    \includegraphics[width=0.48\textwidth]{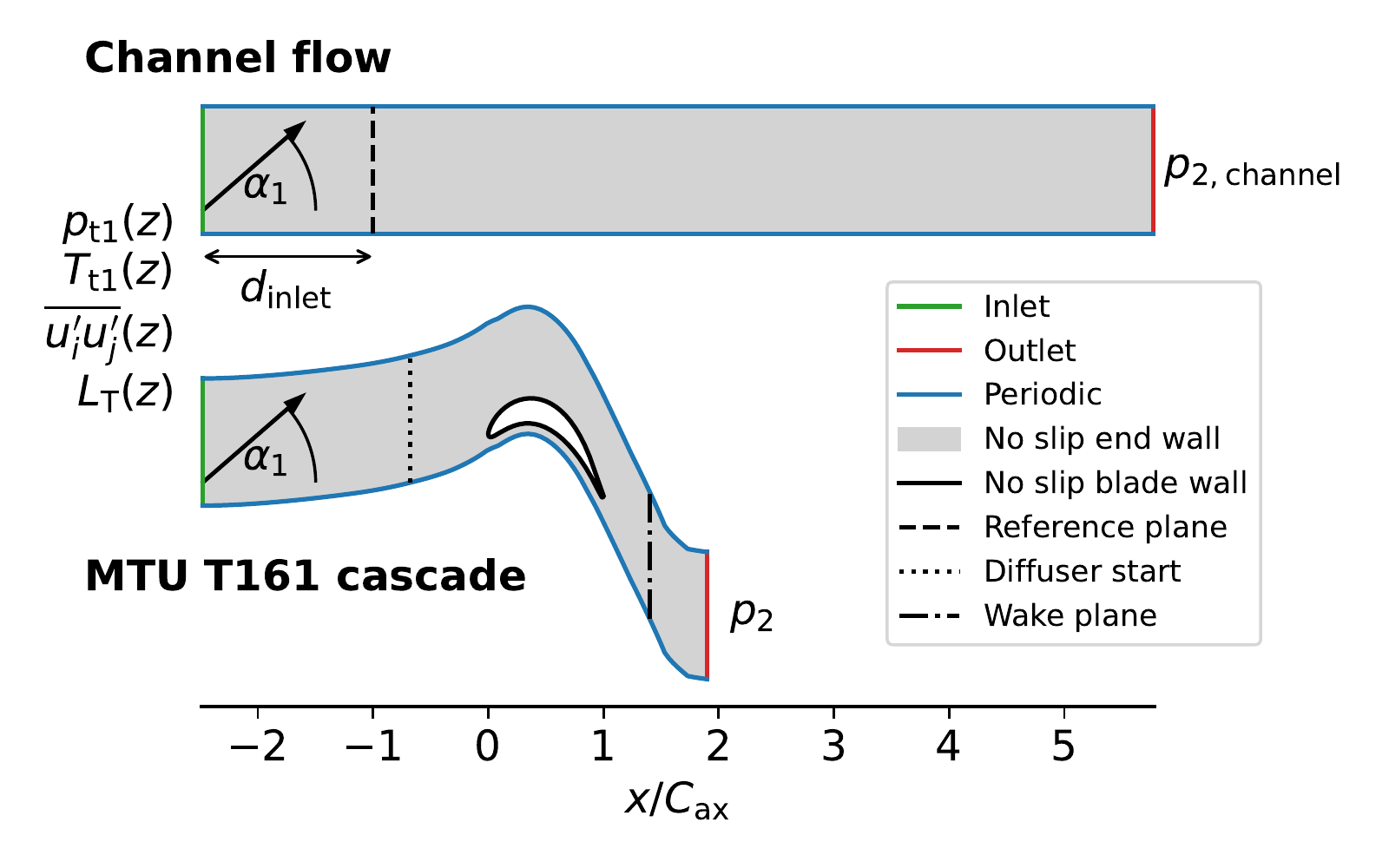}
    \caption{Sketch of the domain and boundary conditions for the precursor channel flow simulation and the simulation of the MTU T161 cascade}
    \label{fig:sketch_domain}
\end{figure}

In practical \ac{CFD}, a significant part of the overall effort is spent obtaining a consistent setup between different computations and the experimental setup of the same physical configuration.
This comprises the representation of geometrical features, manufacturing tolerances, fluid properties and the boundary conditions.
In this particular case, the measurements of have been obtained by mounting the turbine profile in a wind tunnel facility as a linear cascade with seven blades in order to achieve a near-periodic condition on the centre blade.
The blades with a chord length of $C = 0.069935\,\mathrm{m}$ and an average aspect ratio of 2.65 are staggered at an angle of $61.72^\circ$.
The cascade is arranged with pitch to chord ratio of $l_\mathrm{pitch}/C = 0.956$.
The diverging end walls start at $\xbycax = -0.6786$ (blade \ac{LE} at $x=0$) and extend the upstream channel height of $H = 0.16 \,\mathrm{m}$ at an angle of $12^\circ$.
For the simulations, we simplify the geometry by choosing a subdomain of the whole test rig and model only a single blade with periodic boundary conditions in pitchwise direction as sketched in \figref{fig:sketch_domain}.
Downstream of the blade, the domain extends until the point where the diverging end walls in the experiment end and the flow enters a larger volume.
The upstream length of the domain and turbulent inflow boundary conditions are set up to reproduce the end wall boundary layer state and freestream turbulence decay found in the experiment.
In addition to the simulations of the MTU T161, we therefore conduct finite length channel flow simulations with parallel end walls.
This setup reflects the experiments conducted to determine the inflow conditions where the clean wind tunnel without the cascade was used.
As long as it can be guaranteed in the experiment that the inflow conditions with and without blade are comparable, this approach is very beneficial in terms of computational resources required for preliminary studies because of less severe restrictions on mesh size and time step in the channel flow.

\begin{figure}[t]
    \centering
    \includegraphics[width=0.48\textwidth]{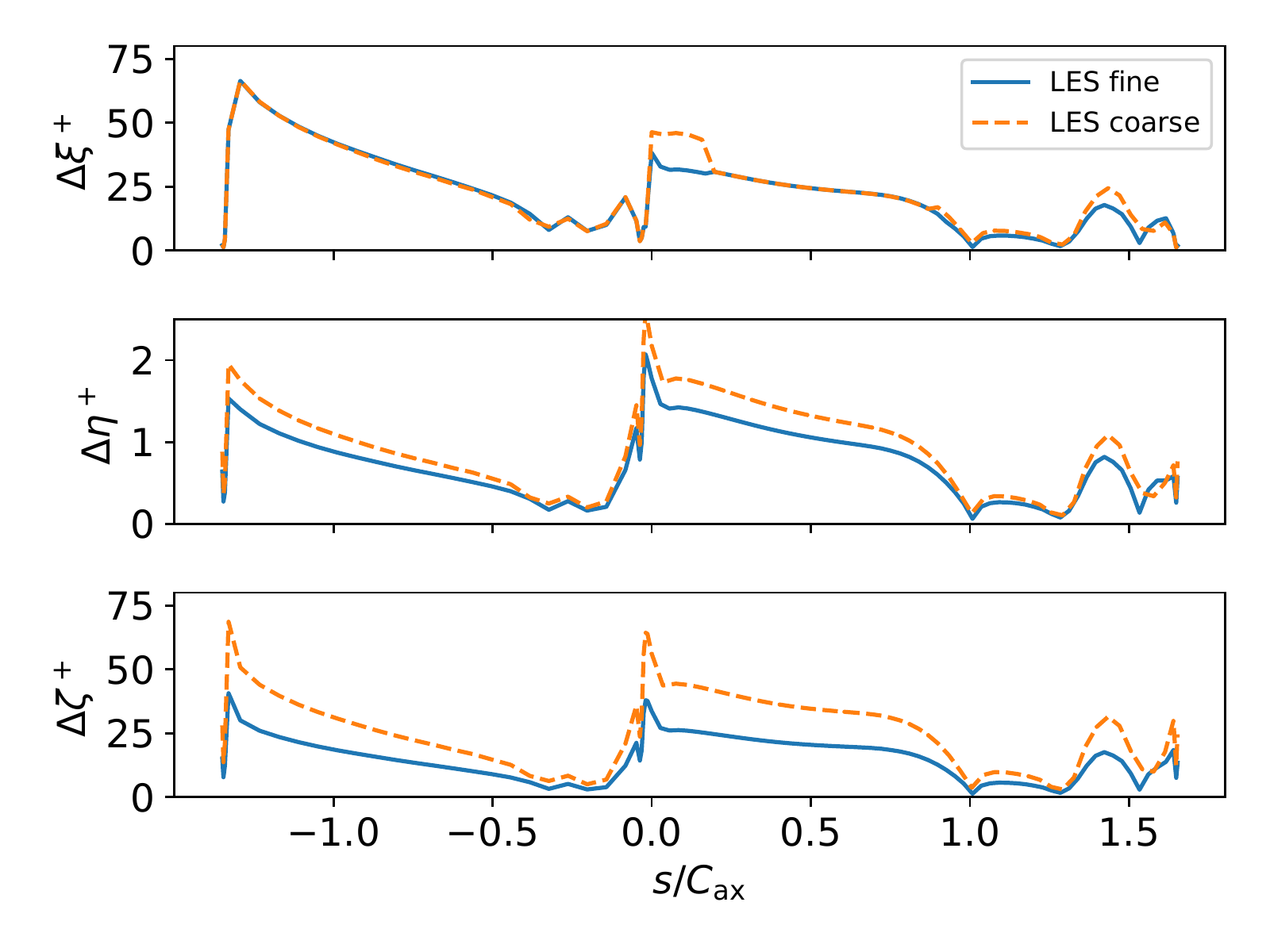}
    \caption{Average solution point distances in streamwise, wall normal and spanwise direction on the blade at midspan (\textit{right})}
    \label{fig:cell_sizes_midspan}
\end{figure}

All \ac{LES} configurations computed for the current paper are summarised in Tab.~\ref{tab:overview-setup}.
The mesh for the turbulent channel flow is constructed as a single block from the rectangular end wall surface with equally sized rectangular faces over the complete domain by an extrusion with the respective wall normal element size stretching.
This results in 83,790 hexahedral elements with geometry polynomial order of $q=4$ to allow for intra-element stretching of solution point distances~\cite{Hindenlang2014}.
For the current flow conditions, we obtain average axial, wall normal and pitchwise element sizes $(\overline{\Delta \xi^+}, \overline{\Delta \eta^+}, \overline{\Delta \zeta^+}) = (32.6, 2.0, 33.1)$ normalised by the solution polynomial order of $N=5$.
For the MTU T161, the mesh is also constructed as extrusion from the end wall surfaces.
These are meshed as a structured multi-block topology with an O-block around the blade to allow for a high quality boundary layer mesh.
Again, we use a polynomial order of $q=4$ to accurately represent the curved geometry of the blade surface.
The O-block is wrapped by a C-block while the rest of the domain is filled with H-blocks putting special focus on the resolution in the blade passage and wake region where the secondary flow structures develop.
The mesh consisting of 876,960 hexahedral elements with a solution polynomial order of $N=5$ (189.4M DoF in total) was designed to meet widely accepted criteria for wall resolution required for \ac{LES}~\cite{Georgiadis2010}.
This is demonstrated in Fig.~\ref{fig:cell_sizes_midspan}, which shows the streamwise, wall normal and spanwise element sizes on the blade surface at midspan with averages of $(\overline{\Delta \xi^+}, \overline{\Delta \eta^+}, \overline{\Delta \zeta^+}) = (23.1, 0.742, 15.0)$, normalised by $N$.
The resolution in the free stream was ensured by the ratio of average solution point distance and estimated Kolmogorov scale below 6 along a mid-passage streamline to meet the requirement of a very well-resolved LES~\cite{Froehlich2005}.
On the end walls, we determine the streamwise and cross-stream directions using the local wall shear stress vector $\mathbf{\tau}_\mathrm{w}$ and the wall normal vector $\mathbf{n}$ as $\mathbf{e}_\parallel = \mathbf{\tau}_\mathrm{w} / \Vert\mathbf{\tau}_\mathrm{w} \Vert$ and $\mathbf{e}_{\perp} = \mathbf{e}_\parallel \times \mathbf{n}$, respectively.
We choose the cell length which is most aligned with the wall shear stress vector as the streamwise cell length $\Delta \xi$ and choose $\Delta \zeta$ accordingly.
Using this procedure, the arithmetic averages are $(\overline{\Delta \xi^+}, \overline{\Delta \eta^+}, \overline{\Delta \zeta^+}) = (23.2, 0.809, 18.8)$ with maximal values of $(\max{\Delta \xi^+}, \max{\Delta \eta^+}, \max{\Delta \zeta^+}) = (65.2, 1.71, 60.3)$.
To be able to assess mesh dependence, results from a preliminary study with a significantly coarser mesh in spanwise direction with 312,424 elements ($q=4$, $N=5$) and 67.9M DoF are also included.

\begin{table}[t]
\caption{Overview of numerical setups. The number of degrees of freedom can be obtained by $\mathrm{nDoF} = n_{xy} \cdot n_z \cdot (N + 1)^3$.}\label{tab:overview-setup}%
\centering
\begin{tabular}{l | r r d{3.1} | rr  | d{2.2} d{3.0}}
\toprule
Configuration & \multicolumn{1}{l}{$n_{xy}$} & \multicolumn{1}{l}{$n_z$} & \multicolumn{1}{l|}{nDoF / $10^6$} & \multicolumn{1}{l}{nMPI} & \multicolumn{1}{l|}{$\mathrm{CPUh}/t_\mathrm{c}$}  & \multicolumn{1}{l}{$\Delta t / 10^{-6}t_\mathrm{c}$} & \multicolumn{1}{l}{$t_{\mathrm{avg}} / t_\mathrm{c}$} \\
\midrule
Channel      & 1710 & 49  &  18.1 &  960 & 131   & 221.9 & 351 \\
Coarse       & 3188 &  98 &  67.9 & 2560 & 5861  & 18.19 & 31  \\
Fine         & 5220 & 168 & 189.4 & 5120 & 25852 & 13.65 & 100 \\
\bottomrule
\end{tabular}
\end{table}

\subsection{Definition of operating point}
For \ac{LPT} cascade cases, the operating point is often specified in terms of the inflow angle $\alpha_1$ and isentropic Mach and Reynolds numbers.
The latter are defined using upstream total pressure $\ptref$ and temperature $\Ttref$, downstream static pressure $\pref$ and a reference chord length $C$ as
\begin{eqnarray}
    \Ma_\isth &=& \sqrt{ \frac{2}{\gamma - 1} \left( \left(\frac{\ptref}{\pref}\right)^{\frac{\gamma - 1}{\gamma}} - 1 \right)}
\\
    \re_\isth &=& \sqrt{\frac{\gamma}{R T_\isth}} \frac{\Ma_\isth \pref C}{\mu_\isth}
\end{eqnarray}
using the ideal gas law with constant $R$, isentropic relations with a coefficient $\gamma$ and Sutherland's law for the viscosity
\begin{eqnarray}
    \mu_{\isth} &=& \mu_\mathrm{ref} \frac{T_\mathrm{ref} + S}{T_\isth + S} \left(\frac{T_\isth}{T_\mathrm{ref}}\right)^{\frac 3 2}
\\
    T_{\isth} &=& \Ttref \left( 1 + \frac{\gamma -1 }{2} \Ma_\isth^2 \right)^{-1}
\end{eqnarray}
with Sutherland's constants $\mu_\mathrm{ref}$, $T_\mathrm{ref}$ and $S$ for air.
The reference quantities are taken to be pitch-averaged centerline conditions.
For the current case, the nominal values are $\alpha_1 = 41^\circ$, $\ptref = 11636 \, \mathrm{Pa}$, $\Ttref = 303.25 \, \mathrm{K}$ at a Reynolds number of $\re_\isth = 90000$ and a Mach number of $\Ma_\isth = 0.6$.
The resulting static outflow pressure is $\pref = 9122.7 \, \mathrm{Pa}$.
These conditions can be easily matched if the flow solver allows to specify the inflow boundary conditions in terms of total pressure, total temperature and flow angle as usually done in turbomachinery \ac{CFD}.
Complications can arise, however, at the intersection of inflow panels and solid walls.
Here, the velocity needs to vanish due to the no-slip condition, which could be easily guaranteed for if the inflow condition was specified using the velocity itself.
If the total pressure is used instead, the inflow velocity close to the wall is a result of the specified total pressure and the static pressure within the domain, which itself is only a result of the simulation determined by the specified outflow pressure and flow through the domain.
Therefore, the specification of boundary layer profiles makes the proper setup of flows with end walls considerably more difficult than simply setting of a free-stream state defined by $\alpha_1$, $\ptref$ and $\Ttref$ away from walls, as it may require adaptation while the simulation progresses.

\subsection{Turbulent inflow}
In addition to reproducing the overall operating point, it is required to match the momentum thickness development of the incoming end wall boundary layers as well as the decay of freestream turbulence given by measurement data at a number of axial stations upstream of the blade.
Since the measurement data is not sufficient to be used as inflow data directly and, more importantly, since synthetically generated turbulence does require a certain adaptation length to develop~\cite{Keating2004}, an indirect approach has to be chosen.

We perform a precursor channel flow simulation to determine the distance from the inlet $d_{\mathrm{inlet}}$ required for the boundary layer to develop and meet the specifications in the reference plane as sketched in \figref{fig:sketch_domain} (arbitrary position due to confidentiality).
The boundary conditions are then set by spanwise varying profiles of $p_{\mathrm{t}, 1}(z)$ and $T_{\mathrm{t}, 1}(z)$ with a constant $\alpha_1$, as well as a spanwise varying Reynolds stress tensor $\overline{u_i' u_j'}(z)$ and turbulent length scale $L_\mathrm{T}(z)$.
At midspan ($z = 0$), the profiles must yield the reference values defining the operating point as described above.
All these profiles will be obtained by scaling a boundary layer profile in wall units at $\re_\mathrm{\theta} = 670$ from \ac{DNS} data (\url{https://www.mech.kth.se/~pschlatt/DATA/}~\cite{Schlatter2010}) and combining it with freestream turbulence values.

Starting from the above centerline inflow conditions, we use the isentropic relations
\begin{equation}
    \frac{T_\mathrm{t}(z)}{T_{1}} = \left( \frac{p_\mathrm{t}(z)}{p_{1}} \right)^{\frac{\gamma-1}{\gamma}} = 1 + \frac{\gamma - 1}{2} \Ma(z)^2
    \label{eqn:isentropic}
\end{equation}
to obtain a spanwise variation.
Here, we assume a constant static temperature $T_{1}$ and pressure $p_{1}$ throughout the boundary layer.
These two quantities can be computed from \eqref{eqn:isentropic} by introducing a centerline Mach number $\Ma_{1,\mathrm{c}}$, which essentially controls the value of the $T_\mathrm{t}$ and $p_\mathrm{t}$ at the end wall.
As discussed above, it has to be chosen such that the total pressure is always greater than the static pressure resulting from the simulation to avoid backflow close to the wall.
An iteration might be necessary to satisfy this condition.
It should be noted that small adaptations of $\Ma_{1,\mathrm{c}}$ mainly influence the total pressure profile close to the wall and do not have a significant impact on the quantities of interest in the simulation.
For this specific case, we chose $\Ma_{1,\mathrm{c}} = 0.362$.
The Mach number at the centerline, however, remains a result of the simulation.

The Mach number profile $\Ma(z)$ is computed based on the discrete \ac{DNS} velocity profile $u_{i,\mathrm{DNS}}^+$ and the corresponding distances to the wall $z^+_{i,\mathrm{DNS}}$ simply using the definitions of $u^+$ and $z^+$:
\begin{eqnarray}
    \Ma(z_i) &=& u^+_{i, \mathrm{DNS}} \cdot \frac{u_\tau}{a_1},
    \\
    z_i &=& z^+_{i,\mathrm{DNS}} \cdot \frac{\nu_1}{u_\tau}
    \label{eqn:wallUnitScaling}
\end{eqnarray}
with
\begin{equation}
    u_\tau = \frac{a_1 \cdot \Ma_{1,\mathrm{c}}}{u_{\infty, \mathrm{DNS}}^+}
\end{equation}
and the DNS freestream velocity $u^+_{\infty, \mathrm{DNS}}$.
The speed of sound $a_1$ and viscosity $\nu_1$ are computed from $T_1$ and $p_1$ using ideal gas and Sutherland laws, respectively.
This allows to describe the spanwise variation of $T_{\mathrm{t}, 1}(z)$ and $p_{\mathrm{t},1}(z)$ using~\eqref{eqn:isentropic}.
The non-dimensional turbulent stress tensor is scaled analogously via
\begin{equation}
\overline{u_j' u_k'}(z_i) = \overline{u_j' u_k'}^+\vert_{i,\mathrm{DNS}} \cdot u_{\tau}^2.
\end{equation}
Finally, the integral turbulent length scale in the boundary layer is estimated from the turbulent kinetic energy and dissipation rate and dimensionalised with $\frac{\nu_1}{u_\tau}$, as it is done in \eqref{eqn:wallUnitScaling}:
\begin{equation}
    L_{\mathrm{T}, \mathrm{BL}}(z_i) = \frac{(k^+_{i, \mathrm{DNS}})^{3/2}}{\epsilon^+_{i, \mathrm{DNS}}} \cdot \frac{\nu_1}{u_\tau}.
\end{equation}

Optimally, the freestream Reynolds stresses and turbulent length scale should be chosen such that the measured turbulent decay is matched.
However, in this case, this would result in a length scale in the order of the cascade pitch.
This could, in principle, be accommodated for by simulating more than one blade at the respective expense of computational effort.
Instead, we chose to decrease the turbulent length scale and scale the non-zero components of the Reynolds stress tensor at the inflow of the domain to reproduce the turbulence intensity in the blade \ac{LE} plane according to the experiment.
This leads to a stronger decay of turbulent structures and has to be considered when assessing the quality of the results.
Turbulence anisotropy, however, is specified as found by hot wire measurements.

The boundary layer profile and freestream values of the turbulence quantities are combined where they intersect at the edge of the boundary layer.
Because of the large freestream turbulent length scale compared to the smaller length scale in the boundary layer, we use
\begin{equation}
    L_\mathrm{T} = \max (L_{\mathrm{T}, \mathrm{BL}}, L_{\mathrm{T}, \mathrm{freestream}})
\end{equation}
for distances to the wall greater than $\delta_{99}$.
Finally, the Reynolds stress tensor is rotated from the streamline-aligned to a Cartesian coordinate system.

For the \ac{RANS} simulations, we use the same boundary layer profile as for the \ac{LES}.
The freestream turbulence, however, is treated differently depending on the model type with the goal of reproducing the turbulence decay of the \ac{LES} and not the experiment.
The turbulence intensity along with a turbulent length scale are specified in the freestream for the \ac{LEVM} setups to determine boundary conditions for the turbulent kinetic energy $k$ and its specific dissipation rate $\omega$.
For the \ac{DRSM} setup, on the other hand, we specify the complete Reynolds stress tensor $\overline{u_i' u_j'}$ and $\omega$ directly to closely match the turbulence anisotropy and decay produced by the \ac{STG}.

\begin{figure}[t]
    \includegraphics[width=0.48\textwidth]{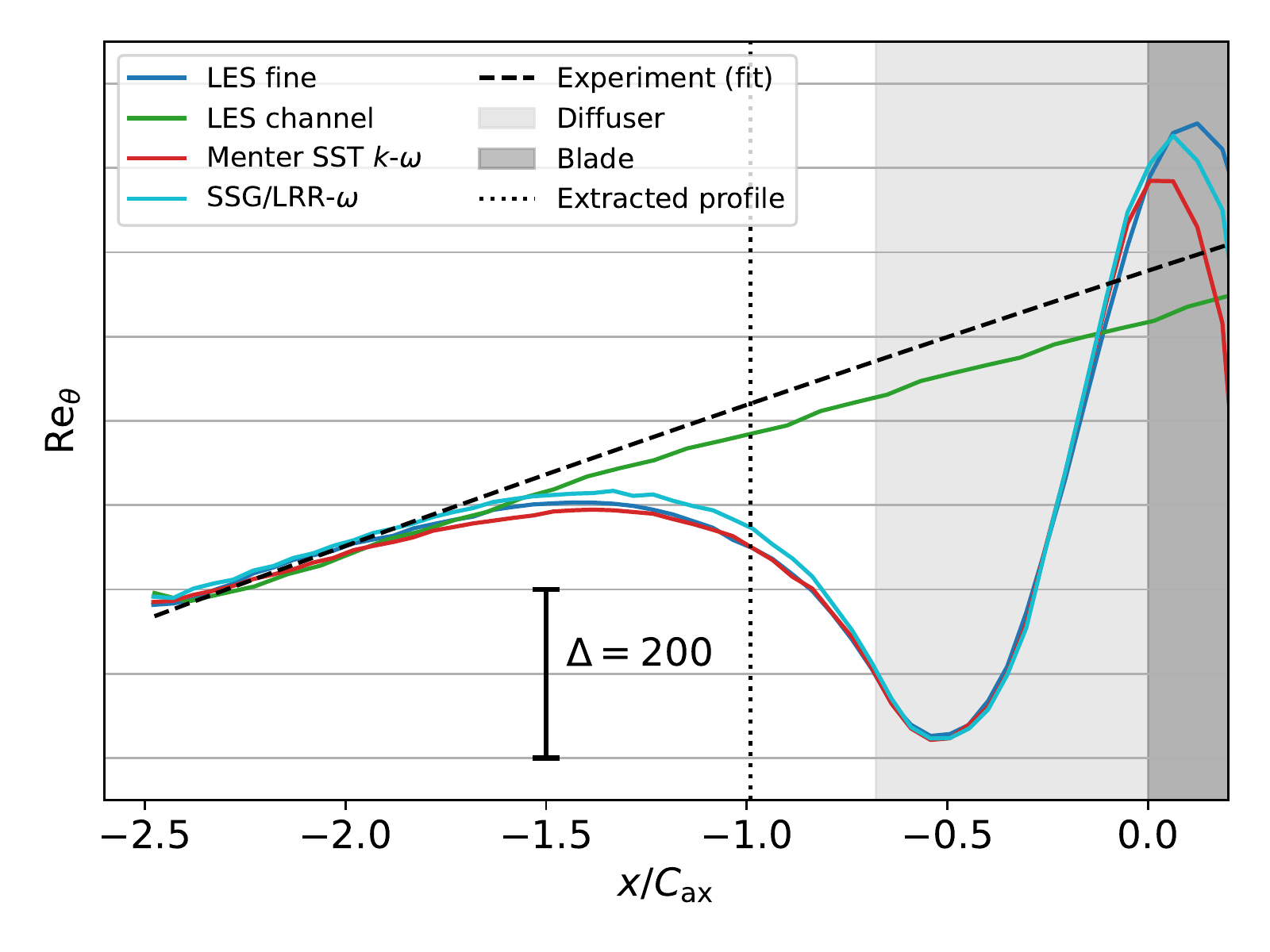}
    \includegraphics[width=0.48\textwidth]{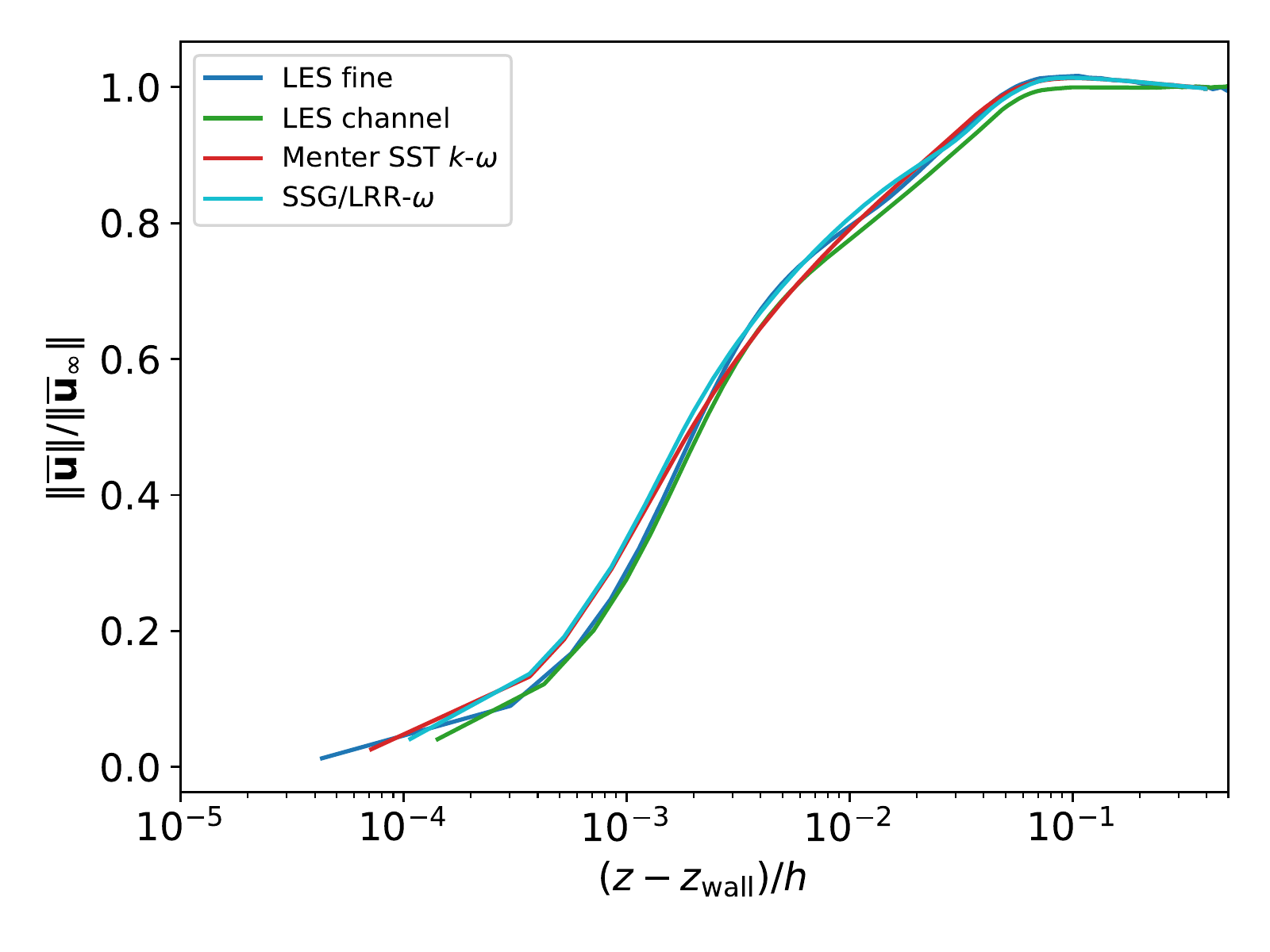}
    \caption{Momentum thickness Reynolds number over axial distance from inlet plane (\textit{left}) and boundary layer velocity profile at $\xbycax = -0.99$ (\textit{right})}
    \label{fig:inflow_boundary_layer}
\end{figure}

In the following we will characterise the inflow boundary layer and freestream turbulence obtained with the approach described above both in terms of the precursor channel flow \ac{LES} and the final simulation with blade on the fine mesh.
Fig.~\ref{fig:inflow_boundary_layer} (\textit{left}) shows the development of the momentum thickness Reynolds number
\begin{equation}
    \re_\mathrm{\theta} = \frac{\Vert\overline{\mathbf{u}}_\infty\Vert \theta}{\nu_\mathrm{wall}},
    \quad
    \theta = \int_0^{\delta_{99}} \frac{\Vert\overline{\mathbf{u}}(z)\Vert}{\Vert\overline{\mathbf{u}}_\infty\Vert} \left( 1 - \frac{\Vert\overline{\mathbf{u}}(z)\Vert}{\Vert\overline{\mathbf{u}}_\infty \Vert} \right) \mathrm dz
\end{equation}
in comparison with a fit of an exponential function proportional to $x^{4/5}$ to the experimental data over the axial distance from the \ac{LE} of the blade $\xbycax$.
We note that the absolute values cannot be shown here due to confidentiality restrictions and that experimental data available for the fit is scarce.
Given these limitations, the precursor channel flow \ac{LES} agrees reasonably well with the experiments.
Because the boundary layer of the simulation with blade (LES fine) is heavily influenced by the diverging end walls and the potential field of the blade, we show the diffusing section of the channel and the extent of the blade as light and dark grey, respectively, in this and all following plots.
Nevertheless, the development of the boundary layer thickness is well matched far upstream from the blade.
Fig.~\ref{fig:inflow_boundary_layer} (\textit{right}) shows normalised boundary layer velocity profiles extracted at $\xbycax = -0.99$ for reference.
The upstream effect of the blade and diffuser is most significant in the wake region of the boundary layer.
Both \ac{RANS} models agree reasonably well with the \ac{LES} in terms of boundary layer shape and integral properties.

\begin{figure}
    \includegraphics[width=0.48\textwidth]{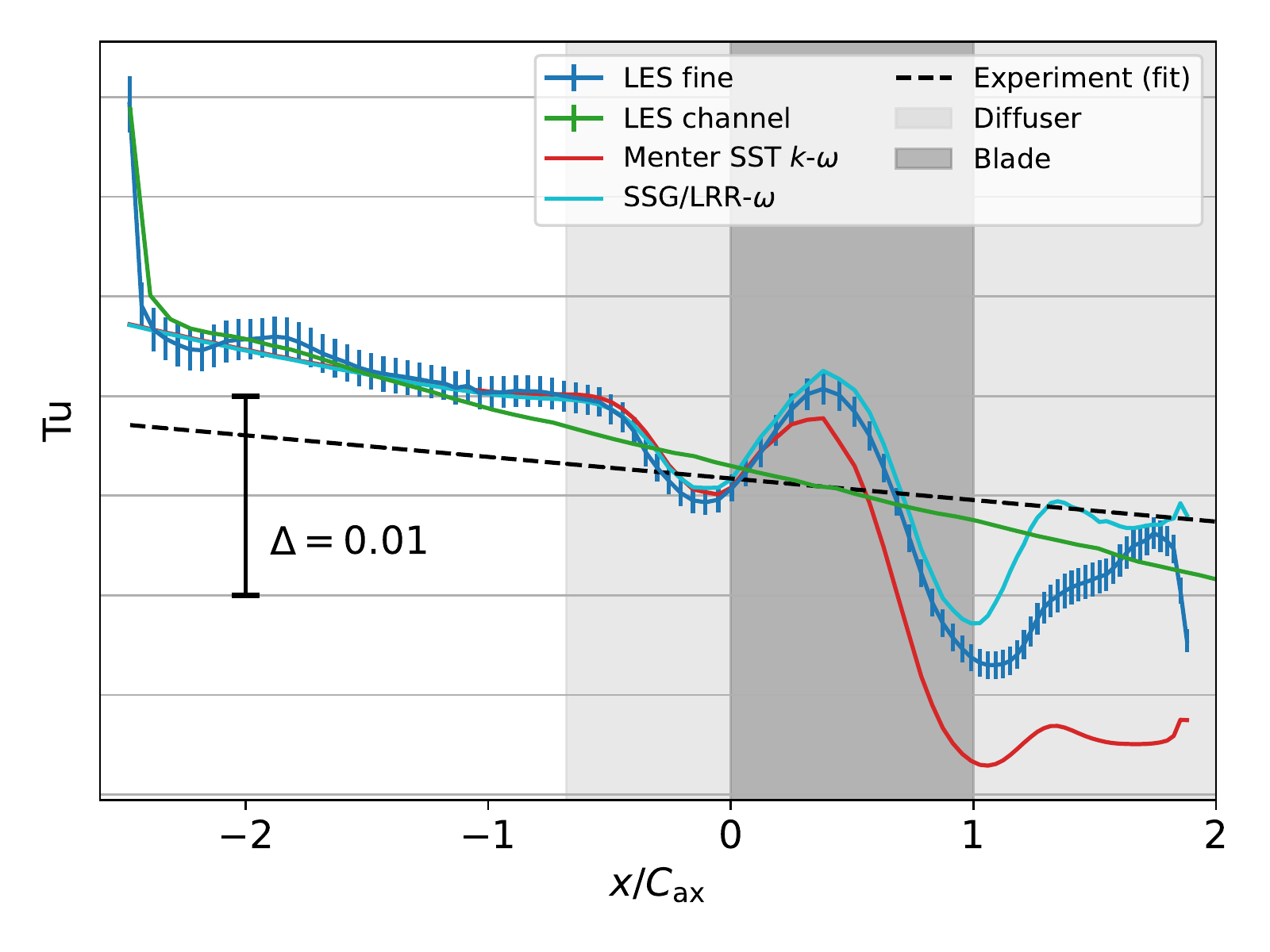}
    \includegraphics[width=0.48\textwidth]{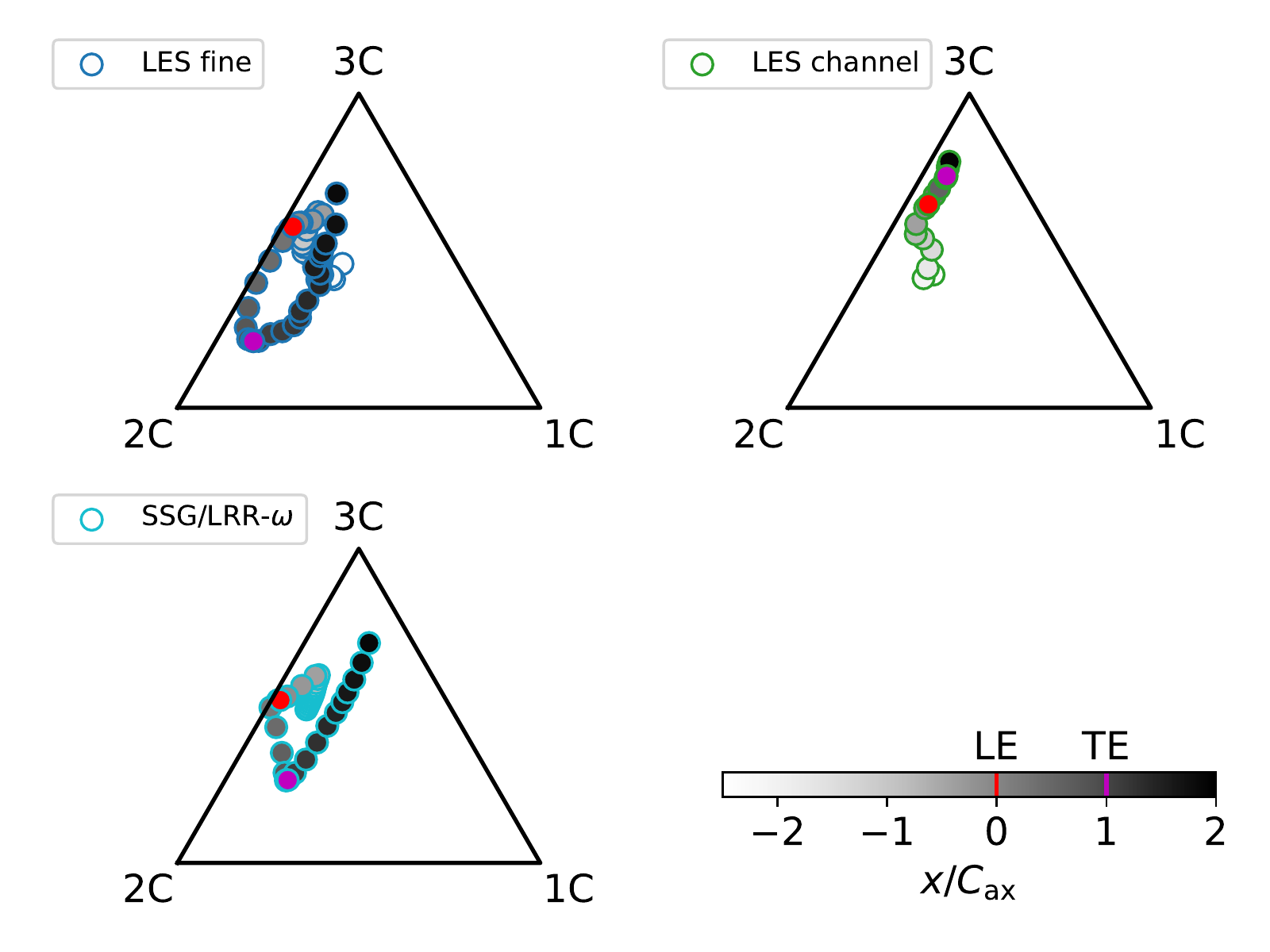}
    \caption{Development of freestream turbulence intensity (\textit{left}) and Reynolds stress anisotropy tensor eigenvalues visualised in barycentric coordinates with the axial position indicated by colour (\textit{right})}
    \label{fig:inflow_intensity_and_anisotropy}
\end{figure}

With the turbulent boundary layer properly defined, we shift our to focus to the characterisation of the freestream turbulence.
The data is obtained from the time-resolved sampling of the flow field along a streamline at midspan passing through the blade \ac{LE} plane at $y / l_\mathrm{pitch} = 0.5$.
Fig.~\ref{fig:inflow_intensity_and_anisotropy} (\textit{left}) shows the development of freestream turbulence intensity $\mathrm{Tu}$ in comparison with a fit to the measured data.
As discussed, the turbulence decay is slightly steeper in our setup but the turbulence intensity at the blade \ac{LE} is well captured.
Compared to the simulation of the MTU T161 (LES fine), a much lower statistical error could be obtained for the precursor channel flow simulation due to less severe time step constraints.
Again, in the upstream area with negligible potential flow effects, both simulations agree very well, before they deviate due to the different geometry.
Both types of \ac{RANS} models are able to closely match the turbulence decay of the \ac{LES}.
Differences start to occur in the blade passage, where the \MenterSST\ model starts to underestimate the turbulence intensity shortly behind the \ac{LE}.
The \SSGLRRw\ model, on the other hand, is able to reproduce the proper turbulence levels throughout the passage before it starts to deviate close to the \ac{TE}.
To achieve this, it was essential to appropriately specify the turbulence anisotropy at the inflow (results with different anisotropy but the same turbulence intensity not shown here).

Fig.~\ref{fig:inflow_intensity_and_anisotropy} (\textit{right}) shows the Reynolds stress anisotropy in terms of the barycentric map~\cite{Banerjee2007} on the same streamline.
Again, both the channel flow and the MTU T161 (LES fine) simulations are shown and the axial position is colour-coded in grey scale with the \ac{LE} marked in red and the \ac{TE} marked in purple.
At the inflow, the injected turbulence has a two-component character (as specified as input to the \ac{STG}) and, after a short length of adaptation, the channel flow shows a return-to-isotropy trajectory (towards the three-component corner).
For the simulation with blade, the effect of flow acceleration on the turbulence structure can be observed.
After an initial decay towards isotropy, the vortices are stretched by the contraction in the blade passage.
In the barycentric triangle, we can observe a trend towards the two-component axisymmetric corner (2C) from \ac{LE} to \ac{TE} along the disk-like axisymmetric border~\cite{Simonsen2005}.
Upon leaving the blade passage, the turbulence exhibits a trend towards the isotropic state.
The results from the \SSGLRRw\ model confirm that the turbulence anisotropy upstream of the blade is consistent with the \ac{LES}.
Although slight deviations in the trajectory become apparent between \ac{LE} and \ac{TE}, the representation of the freestream turbulence anisotropy is qualitatively matched very well.
Since we can only specify isotropic turbulence for the \MenterSST\ model, we deliberately omit this comparison here.

\begin{figure}
    \includegraphics[width=0.48\textwidth]{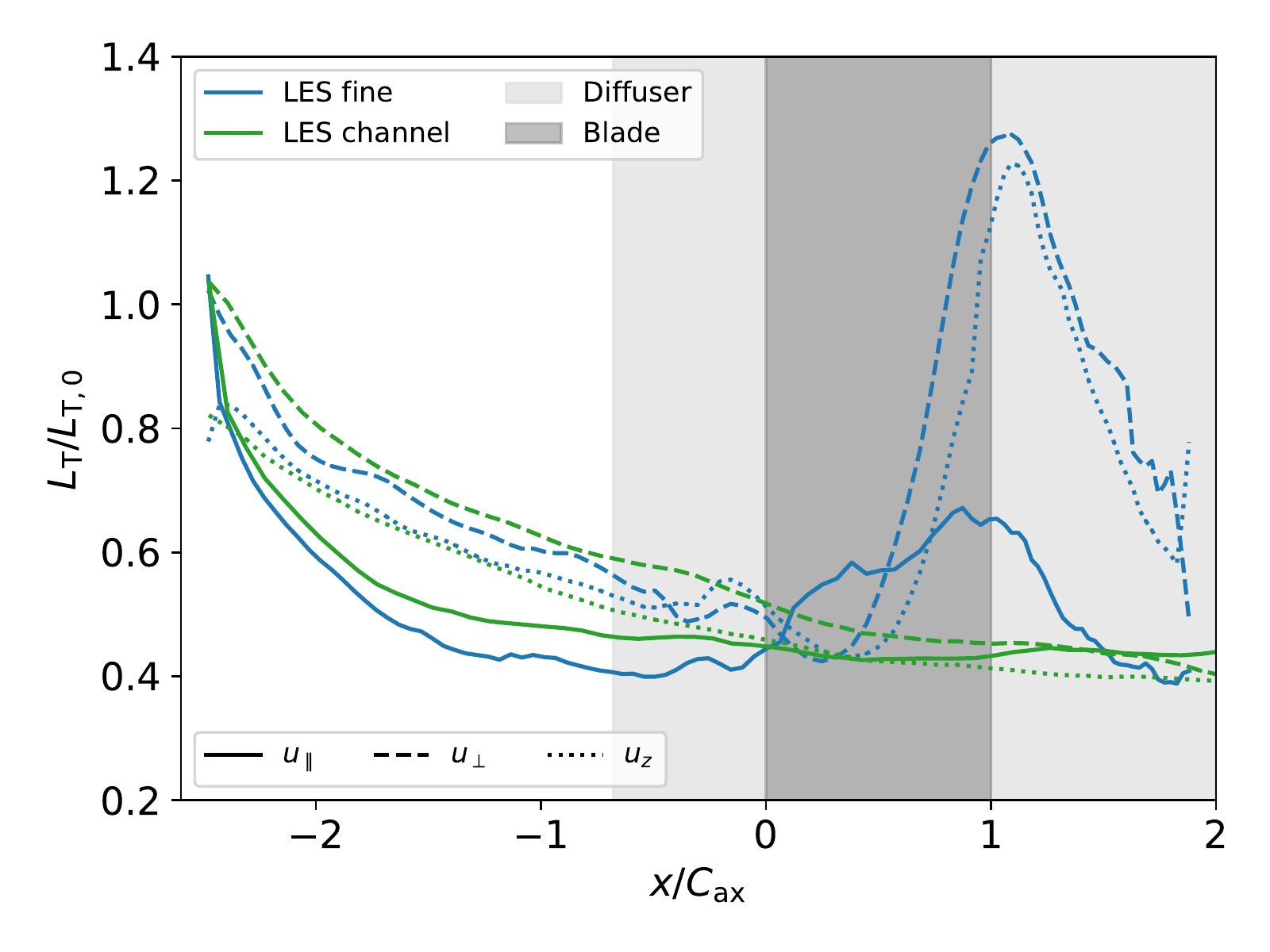}
    \includegraphics[width=0.48\textwidth]{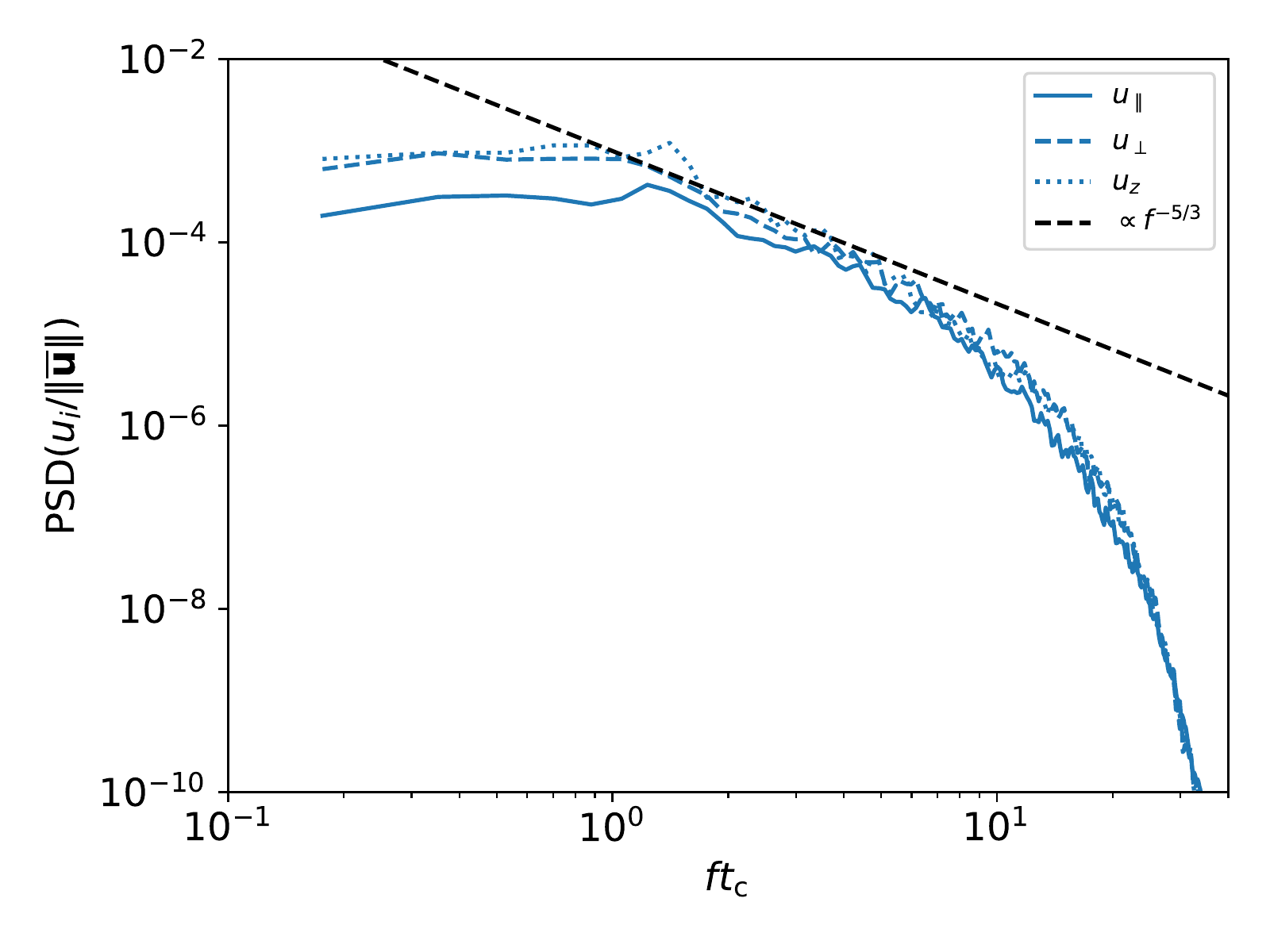}
    \caption{Development of freestream turbulence length scale (\textit{left}) and PSD at $\xbycax = -0.99$ for LES with blade (\textit{right}) for streamwise, cross stream and spanwise velocity components}
    \label{fig:inflow_length_scale_and_PSD}
\end{figure}

Finally, we analyse the spectral content of the streamwise, cross-stream and wall-normal velocity components $(u_\parallel, u_\perp, u_z)$.
The turbulent length scale $L_\mathrm{T}$ normalised with the length scale specified to the \ac{STG} $L_{\mathrm{T}, 0}$ is shown in Fig.~\ref{fig:inflow_length_scale_and_PSD} (\textit{left}).
It has been obtained using the frozen turbulence hypothesis
\begin{equation}
    L_{\mathrm{T}, i} = \Vert\overline{\mathbf{u}}\Vert T_{\mathrm{int}, i}
\end{equation}
with the integral timescale computed as the integral over the normalised auto-correlation of the time signal of the respective velocity component up to the first zero-crossing at $\tau_0$ (no summation over $i$):
\begin{equation}
    T_{\mathrm{int}, i} = \int_0^{\tau_0} \left( \frac{1}{t_\mathrm{avg} \cdot \overline{u_i' u_i'}} \int_0^{t_\mathrm{avg}} u_{i}' (t) u_{i}' (t - \tau) \mathrm dt \right) \mathrm d\tau.
\end{equation}
The axial development shows a relatively rapid decrease in turbulent length scale down to roughly half the specification, which is in line with previous experience with this particular \ac{STG} for decaying isotropic turbulence~\cite{Leyh2020}.
In the blade passage, it can be observed that the length scale in the directions perpendicular to the mean flow directions grows strongly in over the rear part of the blade before the decay towards isotropy sets in downstream.
Fig.~\ref{fig:inflow_length_scale_and_PSD} (\textit{right}) shows the power spectral density in the freestream at $\xbycax = -0.99$.
On the large scales, the turbulence is still considerably anisotropic.
On the smaller scales, however, it has developed an inertial range as indicated by the line proportional to $f^{-5/3}$.

In conclusion, we have presented both an \ac{LES} setup consistent with available experimental data and two \ac{RANS} setups consistent with the \ac{LES}.

\section{Flow analysis}
\subsection{Midspan}
After the characterisation of the incoming flow in the previous section, we now analyse the flow through the MTU T161 cascade with a focus on secondary flow structures and their representation by the \ac{RANS} models.
First, we briefly discuss the consistency between the coarse and the fine mesh LES for the quantities of interest and compare with experiments where applicable to build up trust in the current LES dataset.
After clearing the initial transient, statistics on the fine and coarse mesh were sampled for 100 and 31 convective time units $t_\mathrm{c}$ based on chord length and outflow velocity, respectively, as indicated in Tab.~\ref{tab:overview-setup}.
This difference in number of samples will show in the 68\% confidence intervals~\cite{Bergmann2021}.

\begin{figure}[t]
    \includegraphics[width=0.48\textwidth]{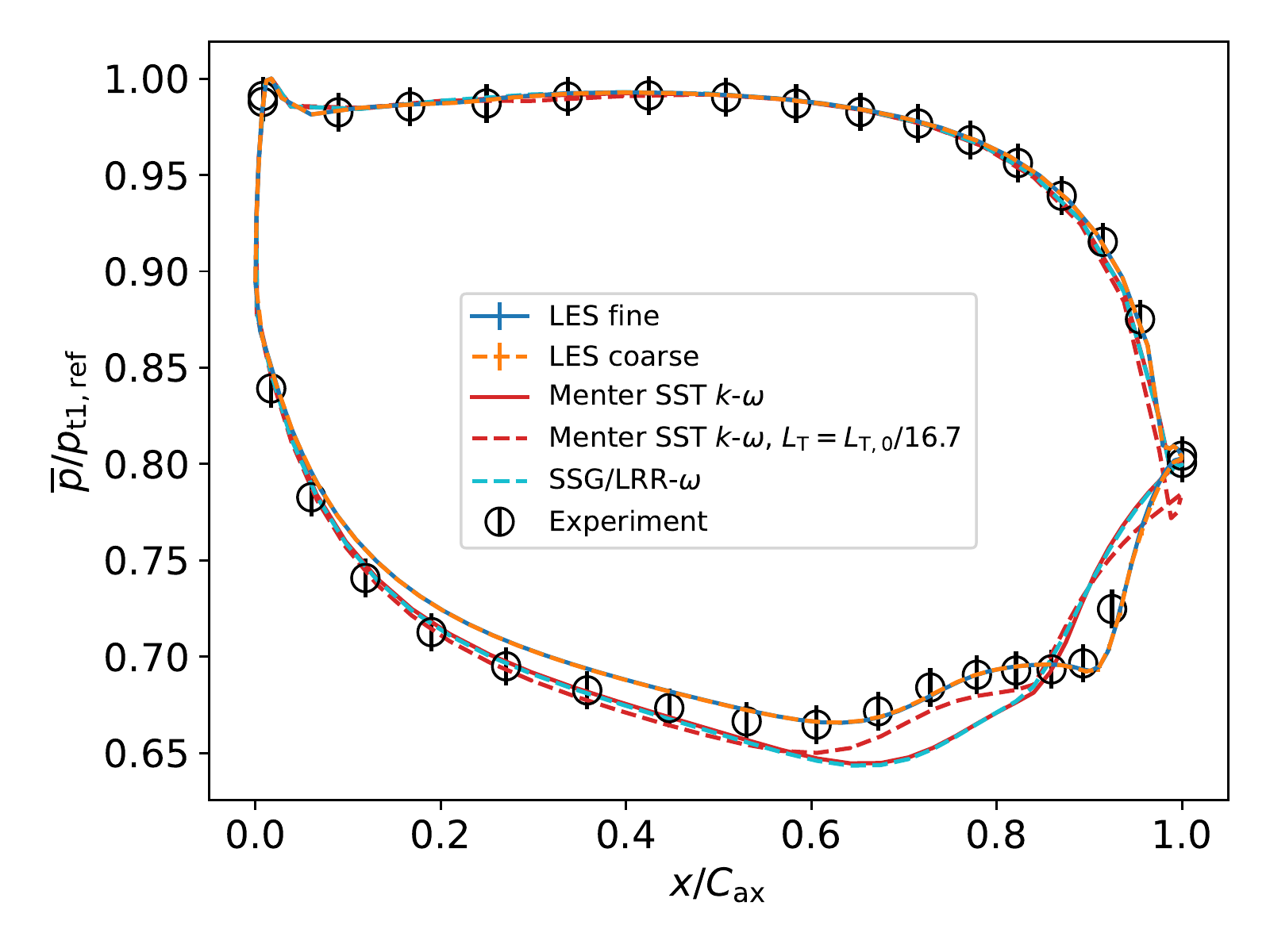}
    \includegraphics[width=0.48\textwidth]{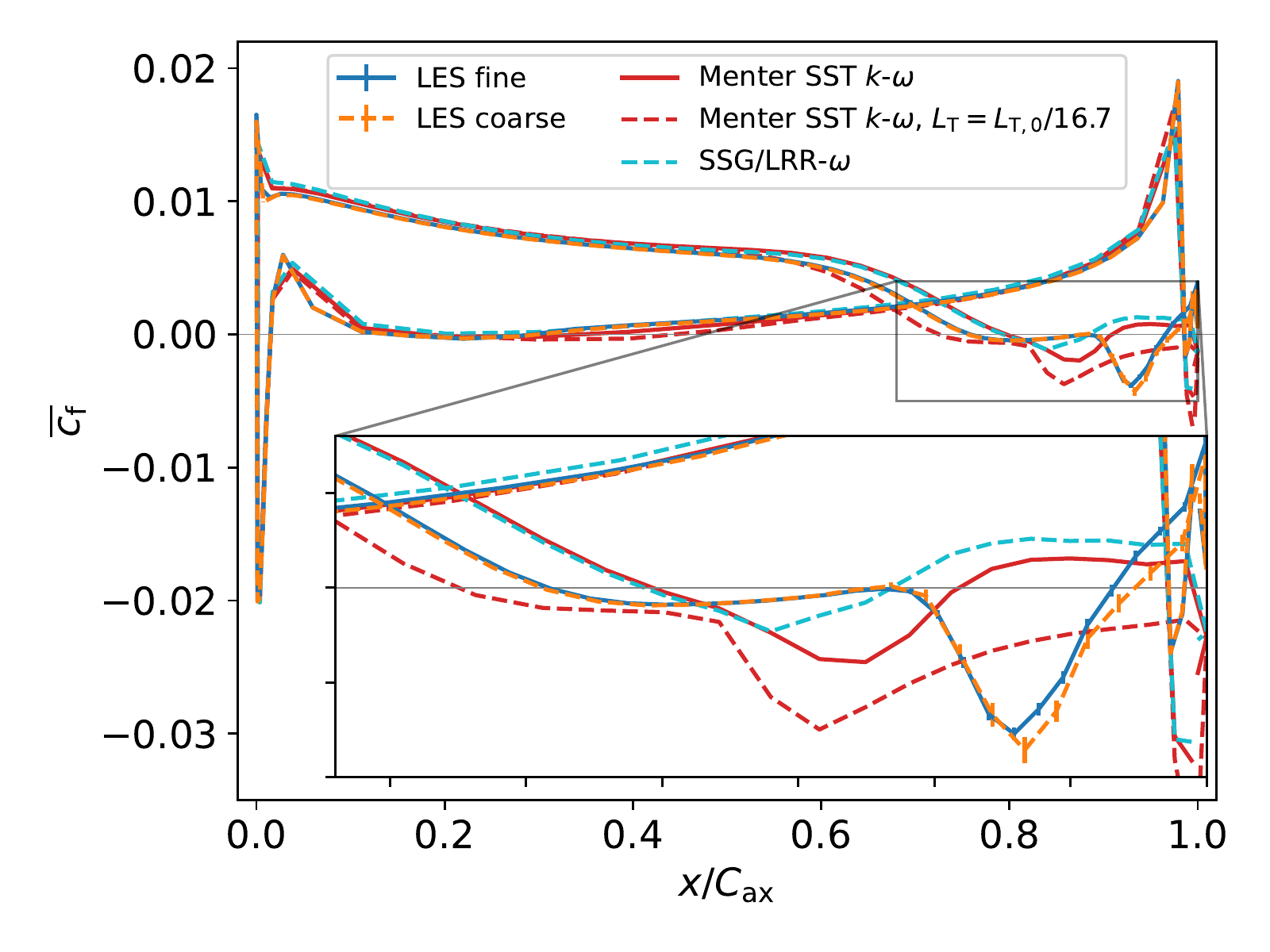}
    \caption{Midspan blade pressure distribution (\textit{left}) and skin friction coefficient (\textit{right}); errorbars indicate 68\% confidence intervals}
    \label{fig:midspan_results}
\end{figure}

\begin{figure}
\centering
    \includegraphicswithtikz[width=0.24\textwidth]{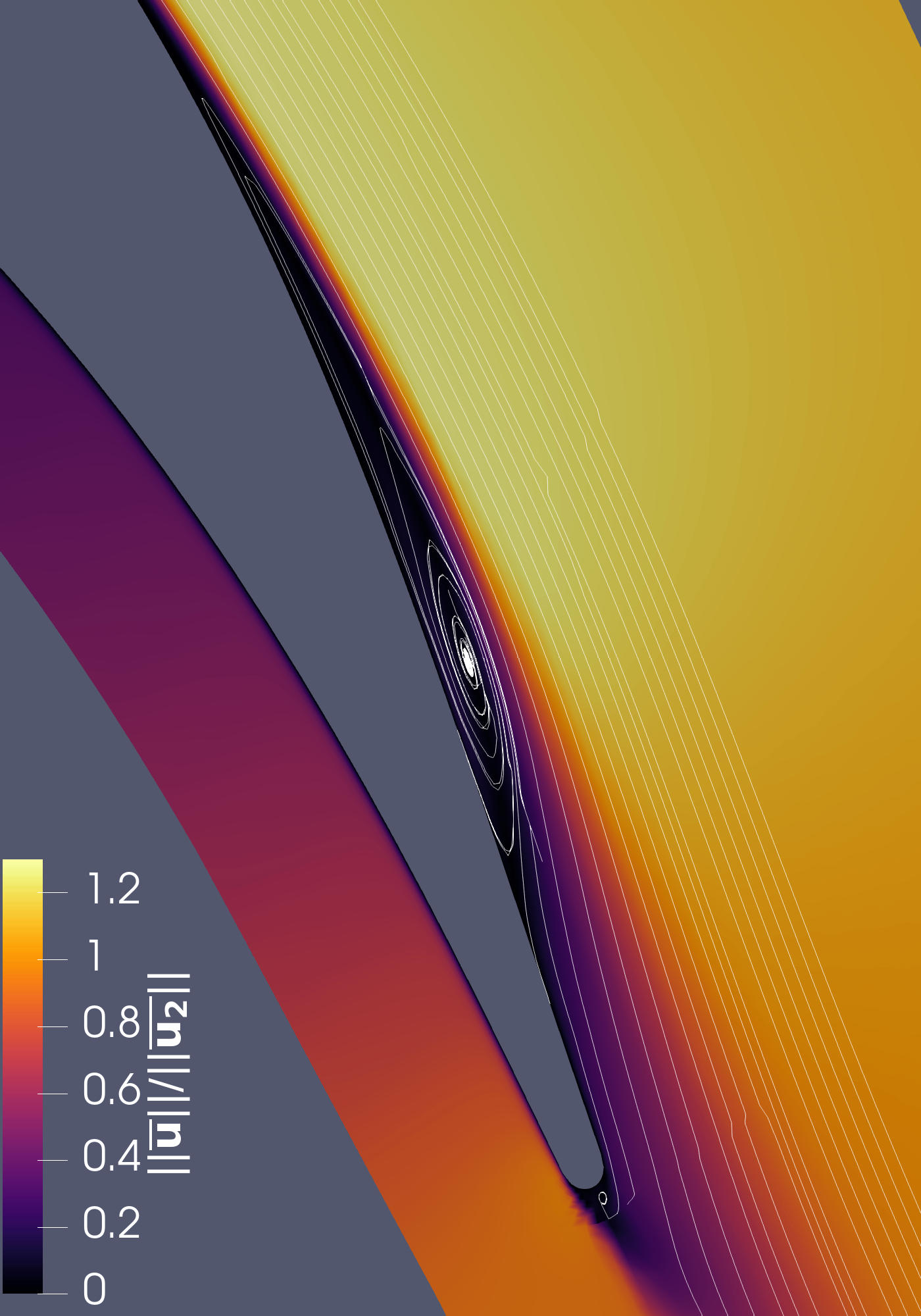}{
        \subfigurelabel{1.4, 2}{LES fine}
    }
    \includegraphicswithtikz[width=0.24\textwidth]{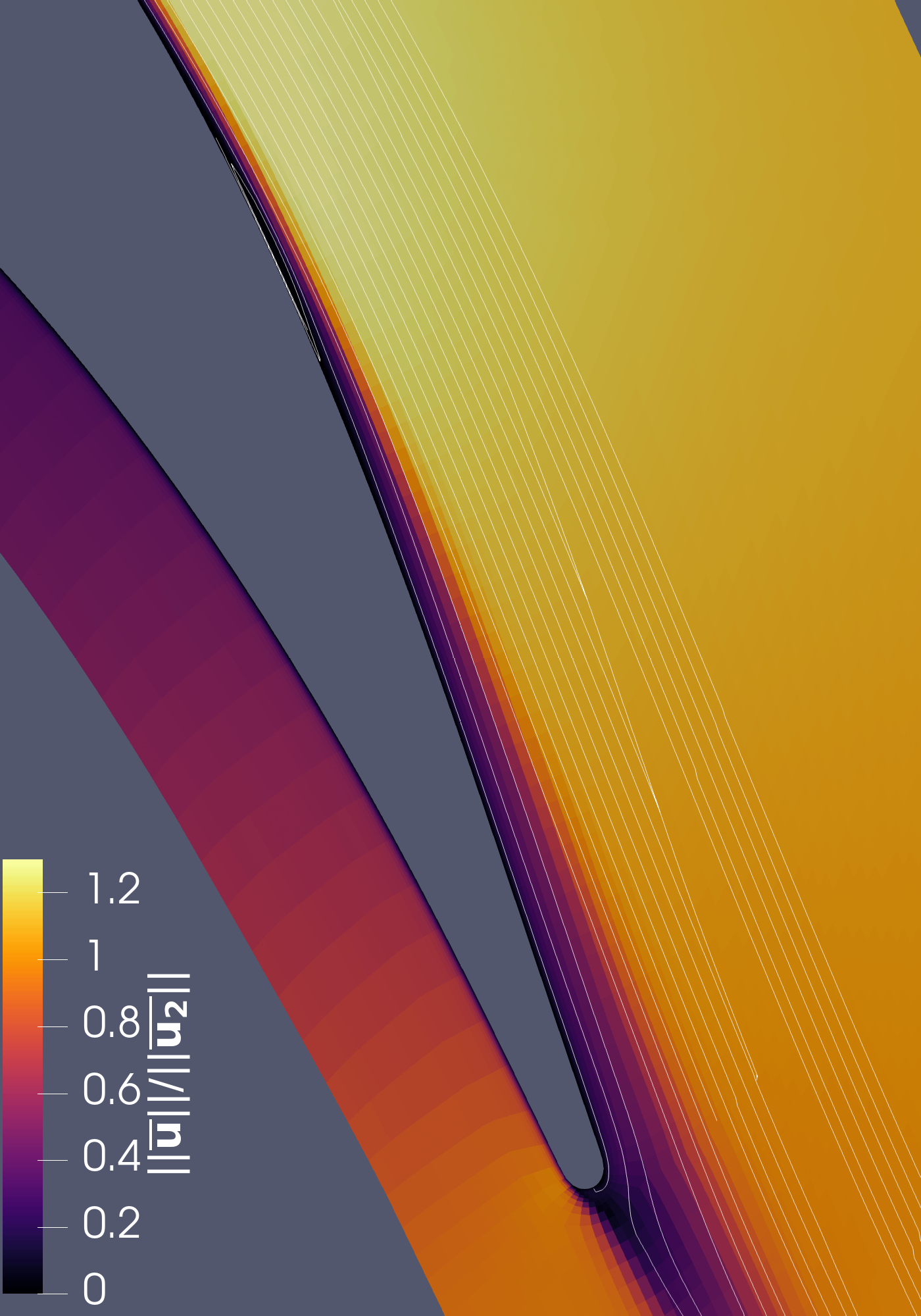}{
        \subfigurelabel{1.4, 2}{\SSGLRRw}
    }
    \includegraphicswithtikz[width=0.24\textwidth]{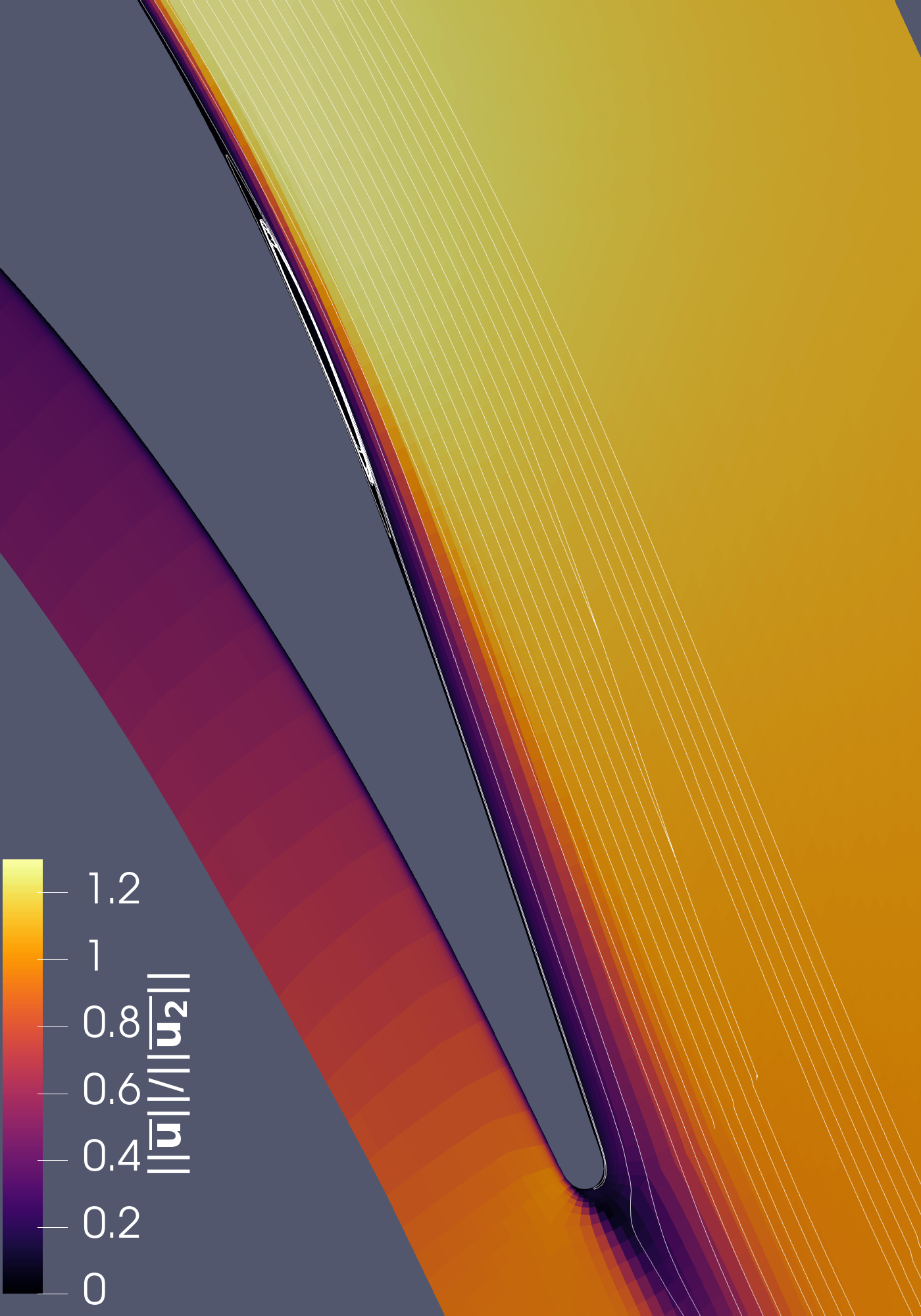}{
        \subfigurelabel{1.4, 2}{\MenterSST}
    }
    \includegraphicswithtikz[width=0.24\textwidth]{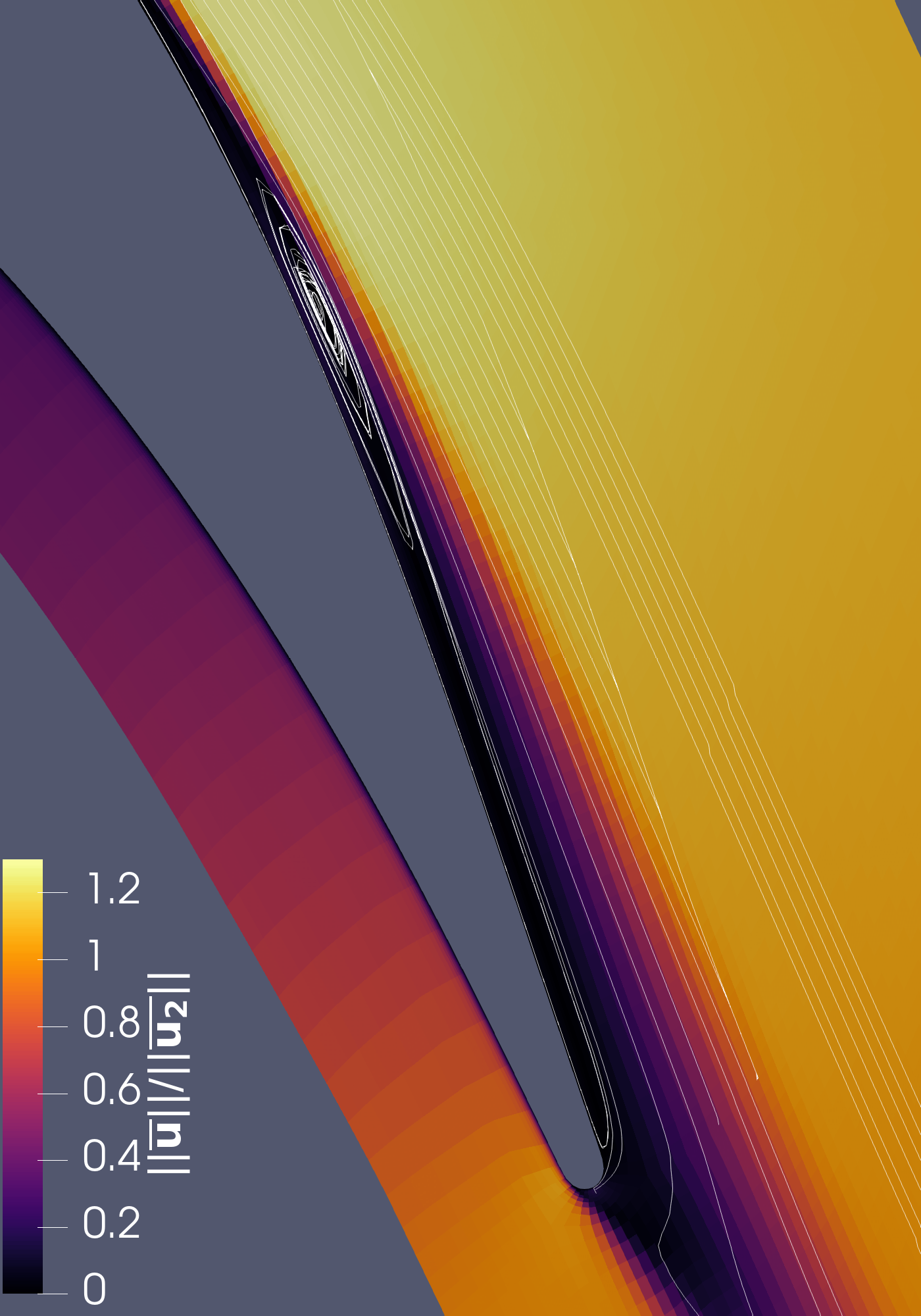}{
        \subfigurelabel{1.4, 2}{\MenterSST\\$L_\mathrm{T} = \frac{L_{\mathrm{T},0}}{16.7}$}
    }
\caption{Midspan streamlines in the separation region of the blade}
\label{fig:midspan_streamlines}
\end{figure}

We start our analysis with the discussion of the blade loading and the wake losses.
Fig.~\ref{fig:midspan_results} shows a comparison of the midspan blade surface pressure distribution with the experiment (\textit{left}) and the skin friction coefficient
\begin{equation}
    \overline{c_\mathrm{f}} = \frac{\mathrm{sgn}(\overline{\tau_{\mathrm{w},x}})\sqrt{\overline{\tau_{\mathrm{w},x}}^2 + \overline{\tau_{\mathrm{w},y}}^2}}{{\ptref} - {\pref}}.
\end{equation}
for which no experimental data are available (\textit{right}).
Both \ac{LES} produce the same blade loading and show only a subtle difference in skin friction downstream of the transition peak of the laminar separation bubble indicating that the resolution on the coarse mesh is not entirely sufficient.
While \ac{LES} and experiment agree very well on the pressure side, a difference in surface pressure can be found on the suction side between $\xbycax = 0.1$ and 0.6.
Similar offsets have been found in previous studies of this configuration, e.g.~\cite{Mueller-Schindewolffs2017}.
The laminar separation bubble and subsequent turbulent reattachment indicated by the pressure plateau between 0.7 and 0.9 and recovery of the base pressure, on the other hand, are captured very well.
Both the \MenterSST\ and the \SSGLRRw\ models fail to replicate this characteristic pressure plateau.
Since it seems to be common practice to compensate for this problem by reducing the inflow turbulent length scale in the \ac{RANS} models, we include such a solution obtained with the \MenterSST\ model as well.
Note, that this leads to a turbulence intensity in the \ac{LE} plane reduced by roughly 56\% compared to the results shown in Fig.~\ref{fig:inflow_intensity_and_anisotropy}.
While this seems to improve the prediction of blade loading at first sight, it can be observed that the \ac{TE} pressure is not recovered, indicating an open separation bubble.
For further interpretation, we evaluate the midspan skin friction coefficient in Fig.~\ref{fig:midspan_results} (\textit{right}) and streamlines in Fig.~\ref{fig:midspan_streamlines}.
The \ac{LES} exhibits the skin friction profile of a classic \ac{LSB} with weak reverse flow after separation followed by a transition peak and subsequent reattachment.
Both \ac{RANS} models with properly configured inflow turbulence levels, however, show a slightly delayed separation and premature reattachment leading to a very thin separation bubble.
The simulation with the adapted turbulent length scale separates early and has the general characteristics of a \ac{LSB} but fails to reattach before the \ac{TE}.
Although this separation bubble is at the wrong streamwise position, its thickness is more comparable to that computed by \ac{LES}.
The larger separation bubble can be explained by the behaviour of the transition model, which keeps a larger part of the boundary layer prior to separation in a laminar state by a value of $\gamma \ll 1$.

\begin{figure}[t]
    \includegraphics[width=0.48\textwidth]{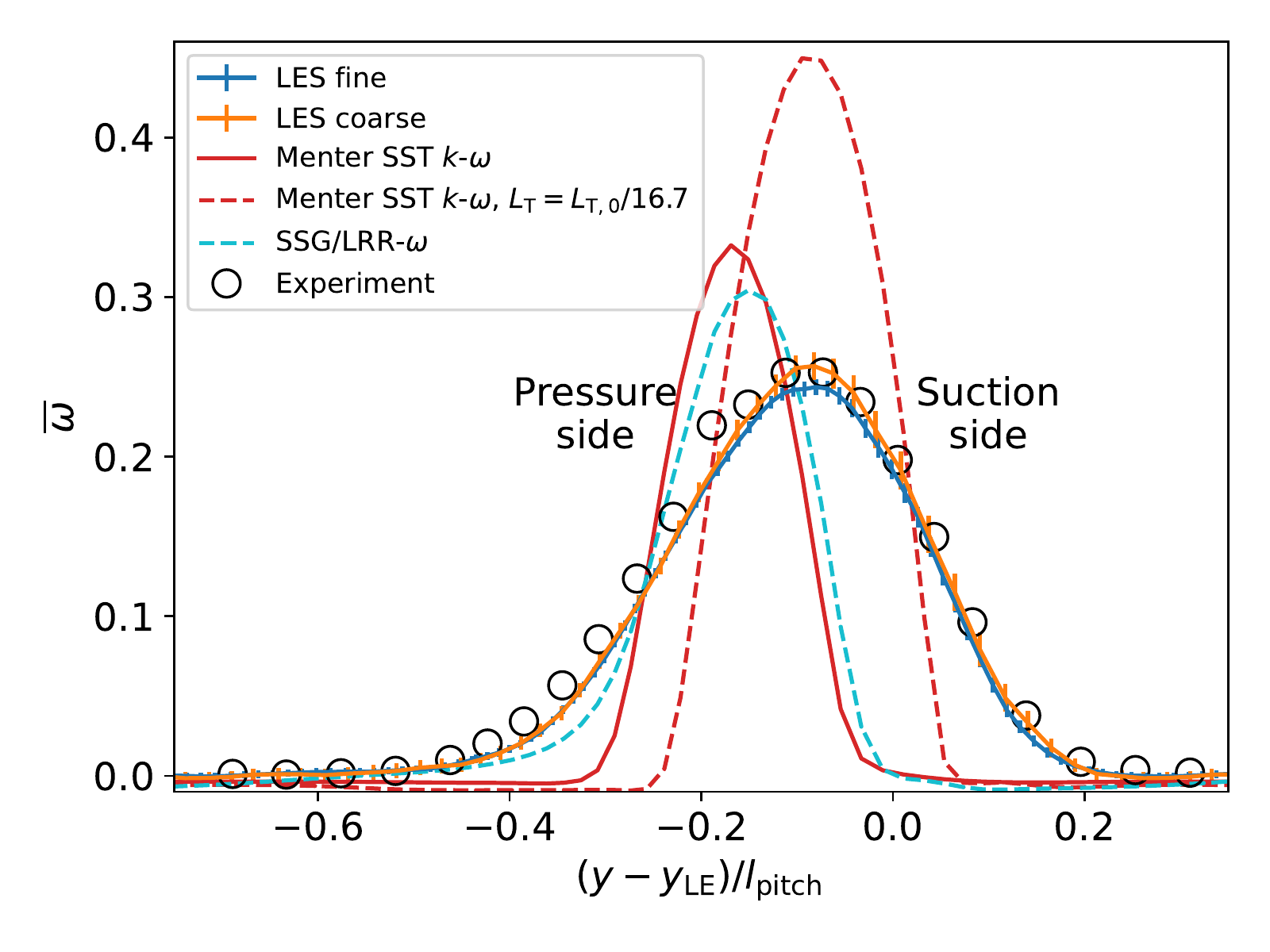}
    \caption{Midspan wake total pressure loss coefficient at $\xbycax = 1.4$}
    \label{fig:wake_midspan}
\end{figure}

Fig.~\ref{fig:wake_midspan} shows the total pressure loss coefficient
\begin{equation}
    \overline{\omega} = \frac{\ptref - p_\mathrm{t}(\overline{\rho}, \overline{u}, \overline{v}, \overline{w}, \overline{p})}{\ptref - \pref},
\end{equation}
computed from time-averaged primitive variables in the wake plane at $\xbycax = 1.4$.
The pitch coordinate $y$ is given with respect to the point $y_\mathrm{LE}$ on the blade at its minimum axial coordinate $\xbycax = 0$.
Both \ac{LES} and the experimental data agree very well within the statistical confidence intervals.
None of the \ac{RANS} setups, on the other hand, is able to reproduce the wake shape accurately.
Due to the very thin separation bubble, both simulations with the appropriate turbulence decay predict a too strong turning of the flow with mass averaged losses underestimated by 43\% and 36\% for the \MenterSST\ and \SSGLRRw, respectively.
The \ac{RANS} simulation with the tweaked turbulent length scale profits from a lucky cancellation of errors and produces a mass averaged loss underestimated by only 12\%.
Similar discrepancies between \ac{RANS} and \ac{LES} in terms of wake development have been reported in the literature for the MTU T161~\cite{Mueller-Schindewolffs2017,Afshar2022} and the T106A~\cite{Marconcini2019}.

\begin{figure}[t]
    \includegraphics[width=\textwidth]{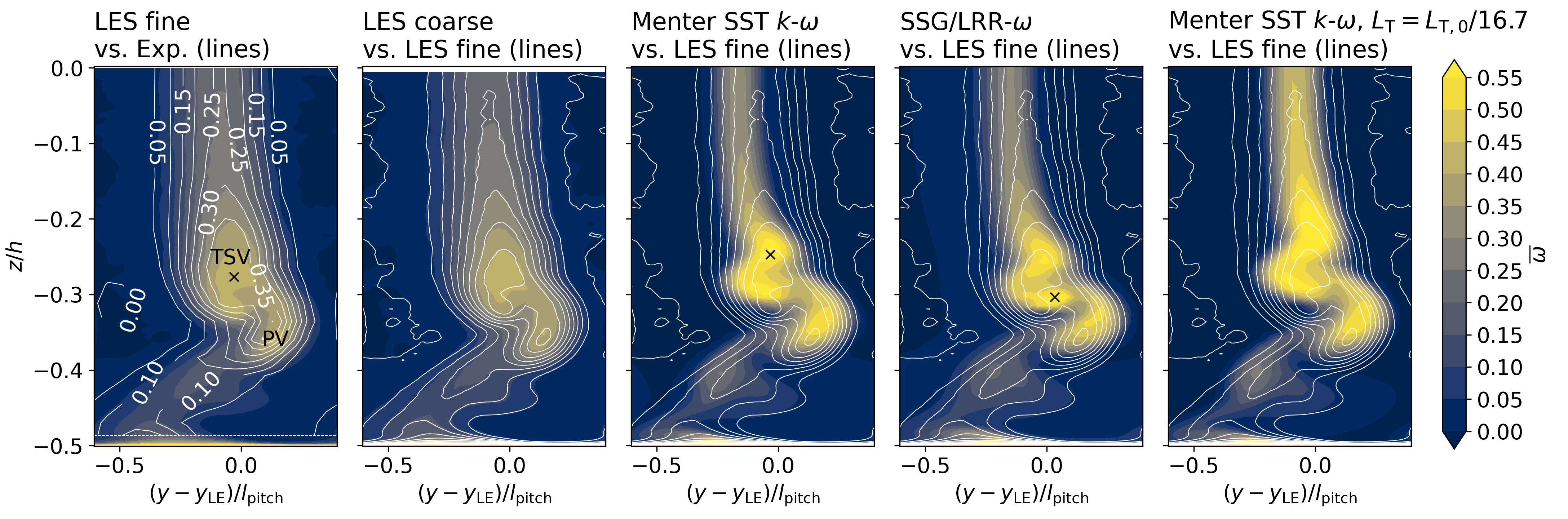}
    \caption{Wake total pressure loss coefficient at $\xbycax = 1.4$}
    \label{fig:wake_2d}
\end{figure}

\subsection{Secondary flow system}
As a final validation of the \ac{LES}, we turn to the secondary flow system, visualised as total pressure loss coefficient in the wake plane at $\xbycax = 1.4$ in Fig.~\ref{fig:wake_2d}.
Note that the $z$-coordinate is given with respect to the local channel height $h$.
The \ac{LES} on the fine mesh is able to reproduce the general shape of the loss cores found in the experiment very well.
The differences in the region of the \ac{TSV} (or counter vortex~\cite{Langston2001,Cui2017}) and \ac{PV} are rather subtle and are visually amplified by the much lower spatial resolution of the experimental dataset.
What can be deduced from the figure is that the loss core associated with the \ac{PV} is predicted slightly closer to the end wall by the current \ac{LES} than in the experiment.
In the second panel of the figure, the \ac{LES} on the coarse mesh is compared to the one on the fine mesh now plotted as contour lines.
As already stated for the midspan results, both simulations give consistent results despite the relatively low spanwise resolution of the coarse \ac{LES}.

The remaining three panels of Fig.~\ref{fig:wake_2d} show the \ac{RANS} results compared with the \ac{LES} on the fine mesh.
A first observation that can be made here is that the tweaking of the freestream turbulent length scale does not significantly modify the secondary flow losses.
Differences between the two \ac{RANS} simulations using the \MenterSST\ model only become apparent around midspan ($z \approx \pm 0.2h$).
We will, therefore, focus on the comparison of the two different \ac{RANS} model approaches (panel 3 and 4) from here on.
The overall shape of the secondary flow loss region depends only weakly on the chosen model and resembles the one obtained with \ac{LES}.
It is noteworthy, however, that \ac{RANS} shows a drastically increased maximum loss coefficient in the region of both the \ac{PV} and the \ac{TSV}.
The position of the maximum is marked by an x in the plot.
With the \MenterSST, it is overestimated by 34\% while the \SSGLRRw\ model results in a 48\% increase with the maximum loss in a different peak compared to both \MenterSST\ and the \ac{LES}.
This is consistent with the results reported by Marconcini et al.~\cite{Marconcini2019} with the difference, that in our case, the \ac{RANS} shows losses that are too large even in a pitch-averaged sense.

\begin{figure}[t]
    \includegraphics[width=0.65\textwidth]{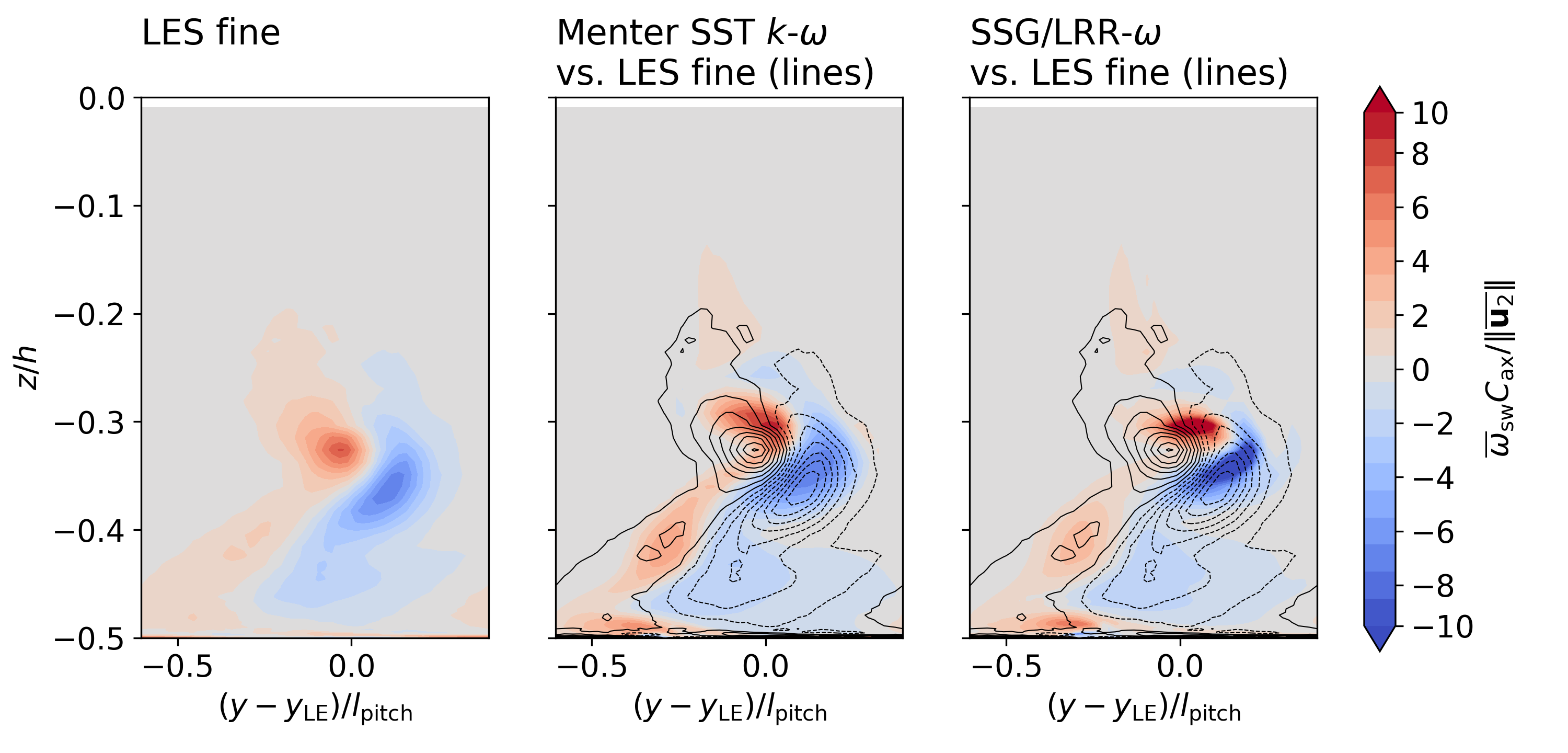}
    \caption{Wake streamwise vorticity $\omega_\mathrm{sw}$ at $\xbycax = 1.4$}
    \label{fig:wake_2d_vorticity}
\end{figure}

An explanation for the increased losses in the \ac{RANS} results is offered by the topology of the streamwise vorticity $\omega_{\mathrm{sw}} = \mathbf{u} \cdot \mathbf{\omega} / \Vert \mathbf{u} \Vert$ shown in Fig.~\ref{fig:wake_2d_vorticity}.
The plots are in the same plane at $\xbycax = 1.4$ as the ones in Fig.~\ref{fig:wake_2d}.
The \ac{RANS} models show increased streamwise vorticity very close to the end wall, indicating that the corner vortex persists further downstream than computed by the \ac{LES}, cf.~\cite{Marconcini2019}.
Moreover, in both models, the system of \ac{TSV} and \ac{PV} is stronger, located at a greater distance from the end wall and slightly offset towards the suction side (positive $y$).
While \acp{DRSM} are often praised for being less diffusive in the representation of vortices, we  have a strong exaggeration of the compactness and strength of both the \ac{TSV} and \ac{PV} by the \SSGLRRw\ model in this case.
To gain further insights into the origin of this discrepancy, we shift our focus further upstream in the paragraphs below.

\begin{figure}[t]
    \includegraphicswithtikz[width=0.64\textwidth]{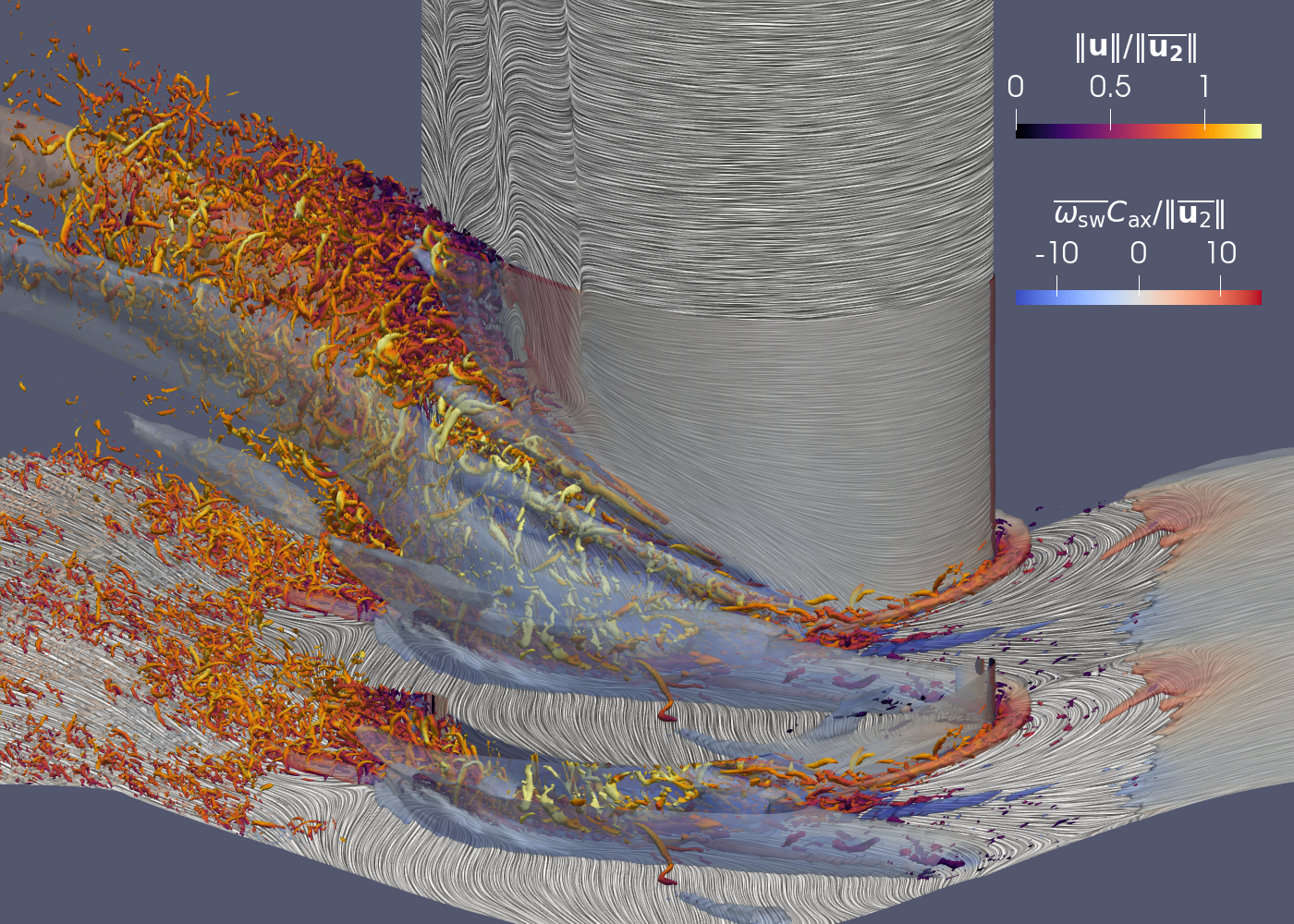}{
        \namednode{0.5, 1.8}{A}{a}
        \simplearrow{a}{-1.0, 1.35}
        \simplelabel{0, -1}{PV}
        \namednode{1, 0.5}{B}{b}
        \simplearrow{b}{-0.4, 0.0}
        \simplearrow{b}{0., -0.5}
        \simplenode{-2.2, -1.3}{C}
    }
    \includegraphicswithtikz[width=0.32\textwidth]{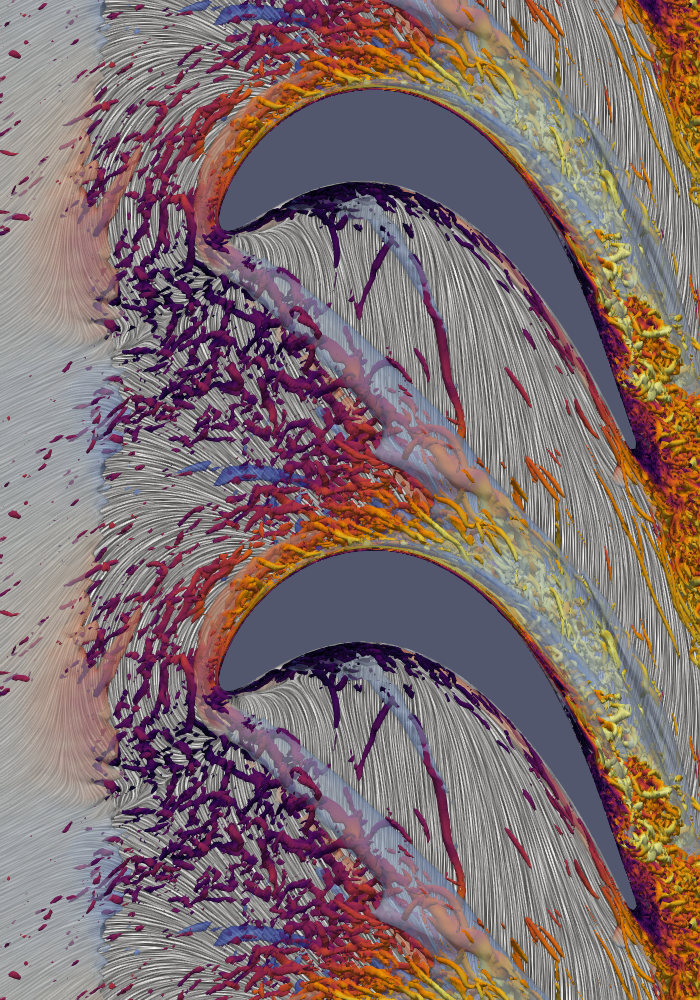}{
        \namedlabel{-1.3, 2}{HSV}{hsv}
        \simplearrow{hsv}{-0.8, 1.5}
        \namedlabel{-1.1, 0}{PV}{pv}
        \simplearrow{pv}{0.2, 0.5}
        \simplenode{-0.8, -2.0}{D}
    }
    \caption{Instantaneous flow field visualised with $Q C_{\mathrm{ax}}^2 / \Vert \overline{\mathbf{u}_2} \Vert^2 = 500$ over the suction side (\textit{left}) and with $Q C_{\mathrm{ax}}^2 / \Vert \overline{\mathbf{u}_2} \Vert^2 = 100$ over the end wall (\textit{left}), both coloured with velocity magnitude; time-averaged solution visualised with surface streaklines and $\overline{Q} C_{\mathrm{ax}}^2 / \Vert \overline{\mathbf{u}_2} \Vert^2 = 1$ iso-surface coloured with streamwise vorticity}
    \label{fig:endwall_flow_instantaneous}
\end{figure}

Fig.~\ref{fig:endwall_flow_instantaneous} provides a three-dimensional view of the secondary flow system in the blade passage near the end walls.
It shows surface streaklines computed with \ac{LIC} using the time-averaged wall shear stress vector and a semi-opaque time-averaged $\overline{Q} C_{\mathrm{ax}}^2 / \Vert \overline{\mathbf{u}_2} \Vert^2 = 1$~\cite{Haller2005} iso-surface coloured in blue-to-red with streamwise vorticity $\omega_\mathrm{sw}$ to indicate the sense of rotation.
Instantaneous vortex structures are visualised as $Q C_{\mathrm{ax}}^2 / \Vert \overline{\mathbf{u}_2} \Vert^2 = 500$ iso-surface over the suction side of the blade (\textit{left}) and as $Q C_{\mathrm{ax}}^2 / \Vert \overline{\mathbf{u}_2} \Vert^2 = 100$ iso-surface in the top view (\textit{right}) allowing to inspect weaker vortex structures of the incoming turbulent boundary layer.
Both are coloured with the same map of velocity magnitude.
The structures of the freestream turbulence are too weak to be visualised by this choice of $Q$.
Towards mid-span, significant turbulent structures are generated over the separation bubble (\circled{A}) and form a major contribution to the wake turbulence downstream.
Increased turbulence activity can also be found within the \ac{PV} and within two separate vortex cores close to the blade surface (\circled{B}), which originate in the same area.
Cui et al.~\cite{Cui2017} also report a separate vortex core above the \ac{PV}, but state that it has an opposite sense of rotation.
In this case, all three vortices rotate in the same direction.
While the end wall boundary layer directly downstream of the \ac{PV} does not yield any significant instantaneous vortical flow structures, strong turbulent flow begins to reappear downstream of a saddle point close to the suction side \ac{TE} (\circled{C}).
The top view of the end wall with the lower value of $Q$ confirms that the incoming turbulent boundary layer (\circled{D}) is completely lifted off the wall by the \ac{PV} resulting in the downstream calmed region, cf.~\cite{Cui2017}.

\begin{figure}[t]
    \includegraphicswithtikz[width=0.32\textwidth]{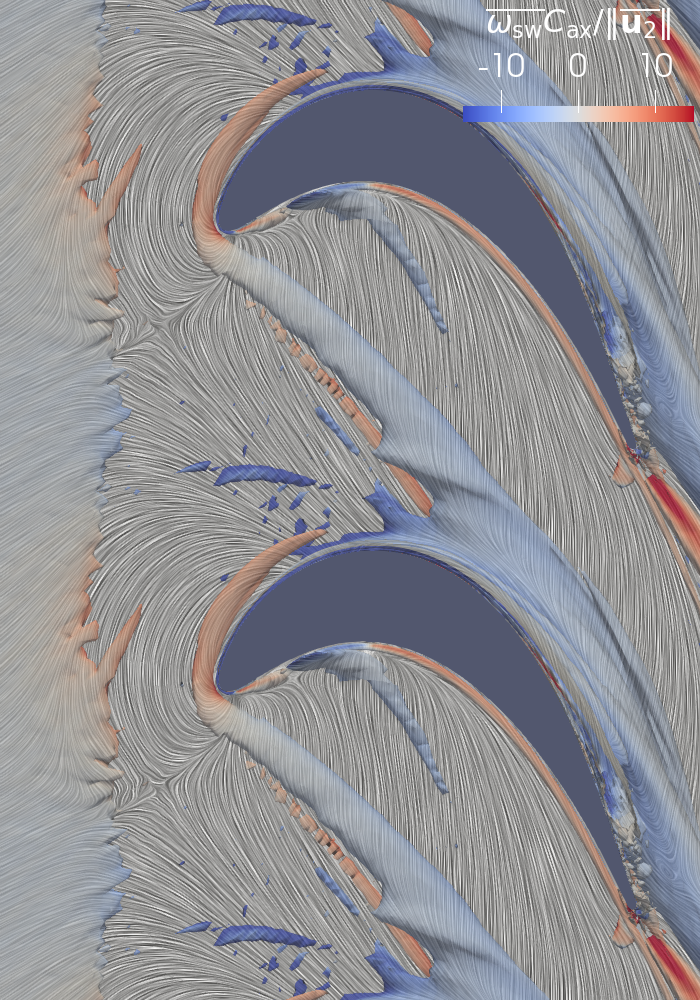}{
        \namedlabel{-1.3, 2}{HSV}{hsv}
        \simplearrow{hsv}{-0.8, 1.5}
        \namedlabel{-1.1, 0}{PV}{pv}
        \simplearrow{pv}{0.2, 0.5}
        \namedlabel{0.8, -2.3}{LECV}{lecv}
        \simplearrow{lecv}{0.6, -1.0}
        \namedlabel{1.5, -1.0}{TSV}{tsv}
        \simplearrow{tsv}{1.7, 0.0}
        \namednode{1, 0.5}{A}{a}
        \simplearrow{a}{0.32, 1.3}
    }
    \includegraphicswithtikz[width=0.32\textwidth]{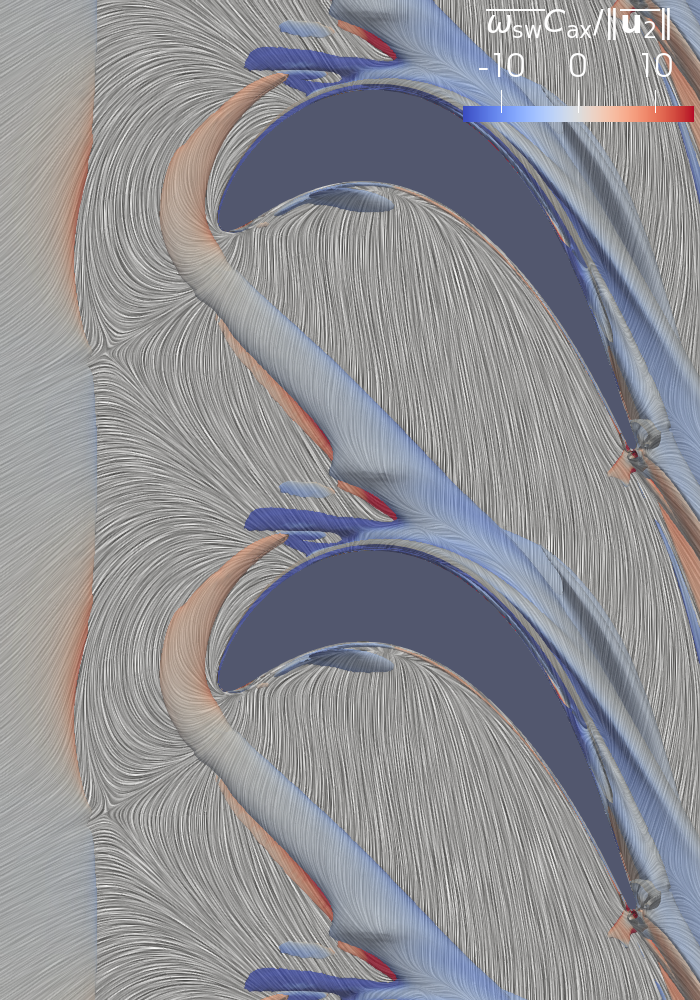}{}
    \includegraphicswithtikz[width=0.32\textwidth]{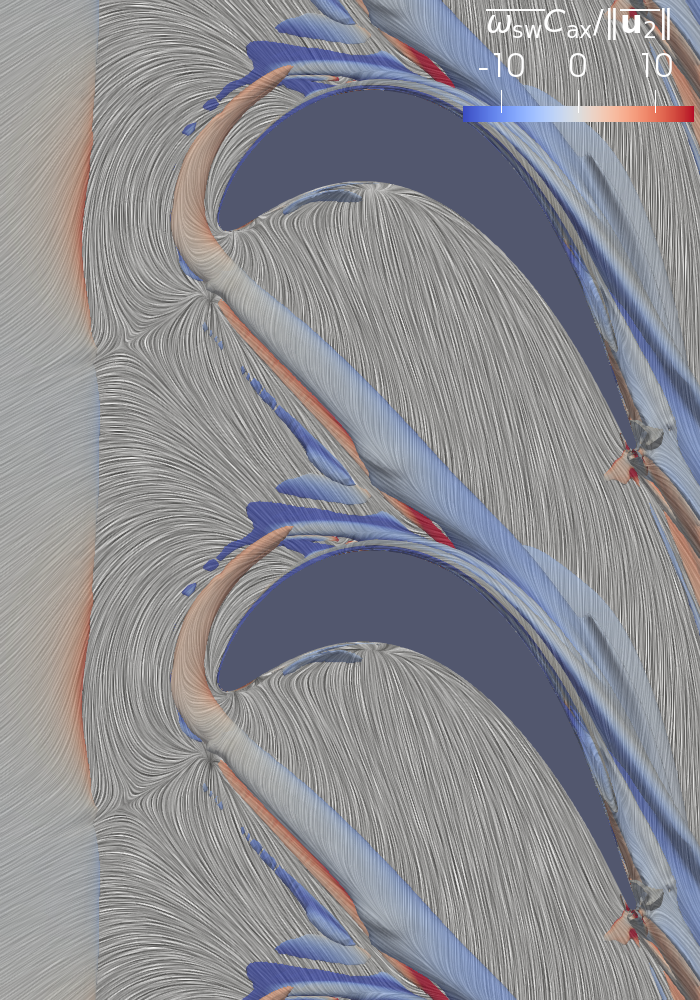}{
        \namedlabel{-1.5, -1}{SP2}{sp2}
        \simplearrow{sp2}{-0.8, -1.4}
        \namedlabel{-1.5, -2.2}{SP1}{sp1}
        \simplearrow{sp1}{-1.2, -1.7}
    }
    \caption{Surface streaklines on the end wall and vortices by $Q$ isosurface coloured with streamwise vorticity for LES average (\textit{left}), \MenterSST\ (\textit{middle}) and \SSGLRRw\ (\textit{right})}
    \label{fig:endwall_flow_vortices}
\end{figure}

A less obstructed view of the averaged flow can be found in Fig.~\ref{fig:endwall_flow_vortices}, which shows the same view as Fig.~\ref{fig:endwall_flow_instantaneous} (\textit{right}), but without the instantaneous structures.
In addition to the \ac{LES} (\textit{left}), the figure shows results for \MenterSST\ (\textit{middle}) and \SSGLRRw\ (\textit{right}) for comparison.
The \ac{HSV} develops due to a roll-up of the incoming boundary layer.
But while its suction side leg (red) dissipates well before the suction peak, its pressure side leg (blue) follows the passage cross flow towards the suction side of the next blade, lifts off the end wall and develops into the \ac{PV}.
Behind the blade, a second large structure can be identified as the \ac{TSV}, which rotates in the opposite direction.
Both vortices persist well downstream of the blade and become visible as the two regions of strong pressure loss in Fig.~\ref{fig:wake_2d} and vorticity in Fig.~\ref{fig:wake_2d_vorticity}.
Another small but very distinct vortex developing along the pressure side of the blade is the \ac{LECV}.
The pressure side features a short separation bubble close to the \ac{LE}.
Due to the diverging end walls, the backflow within the bubble is driven towards the end walls where it rolls up, lifts off and mixes with the newly developing boundary layer in the passage between the \ac{PV} and the pressure side of the blade (\circled{A}).

With the \ac{RANS} models, the general topology of the secondary flow can be reproduced quite reasonably.
The most striking topological difference between the models is the appearance of a second saddle point (SP2) between the \ac{LE} and the first saddle point (SP1) slightly upstream for the \SSGLRRw\ model.
Apart from that, the \ac{DRSM} produces more compact vortex cores than the \ac{LEVM} as already observed in the downstream wake plane.
Both models predict the \ac{HSV} to appear at nearly the same position and follow the same trajectory with the pressure side leg following a nearly straight path.
The \ac{LES}, on the other hand, shows the \ac{HSV} to originate closer to the \ac{LE}.
Its suction side leg follows the blade surface more closely and the pressure side leg curves through the passage before reaching the adjacent blade.
Another difference is the vortex resulting from the \ac{PS} separation (\circled{A}), which is present in the \ac{LES}, but visible only weakly for the \MenterSST\ model and almost impossible to identify for the \SSGLRRw\ model.
This is caused by a weaker \ac{PS} separation, which for the \SSGLRRw\ model, does not even extend over the entire blade span.

\begin{figure}
    \centering
    \includegraphics[width=\textwidth]{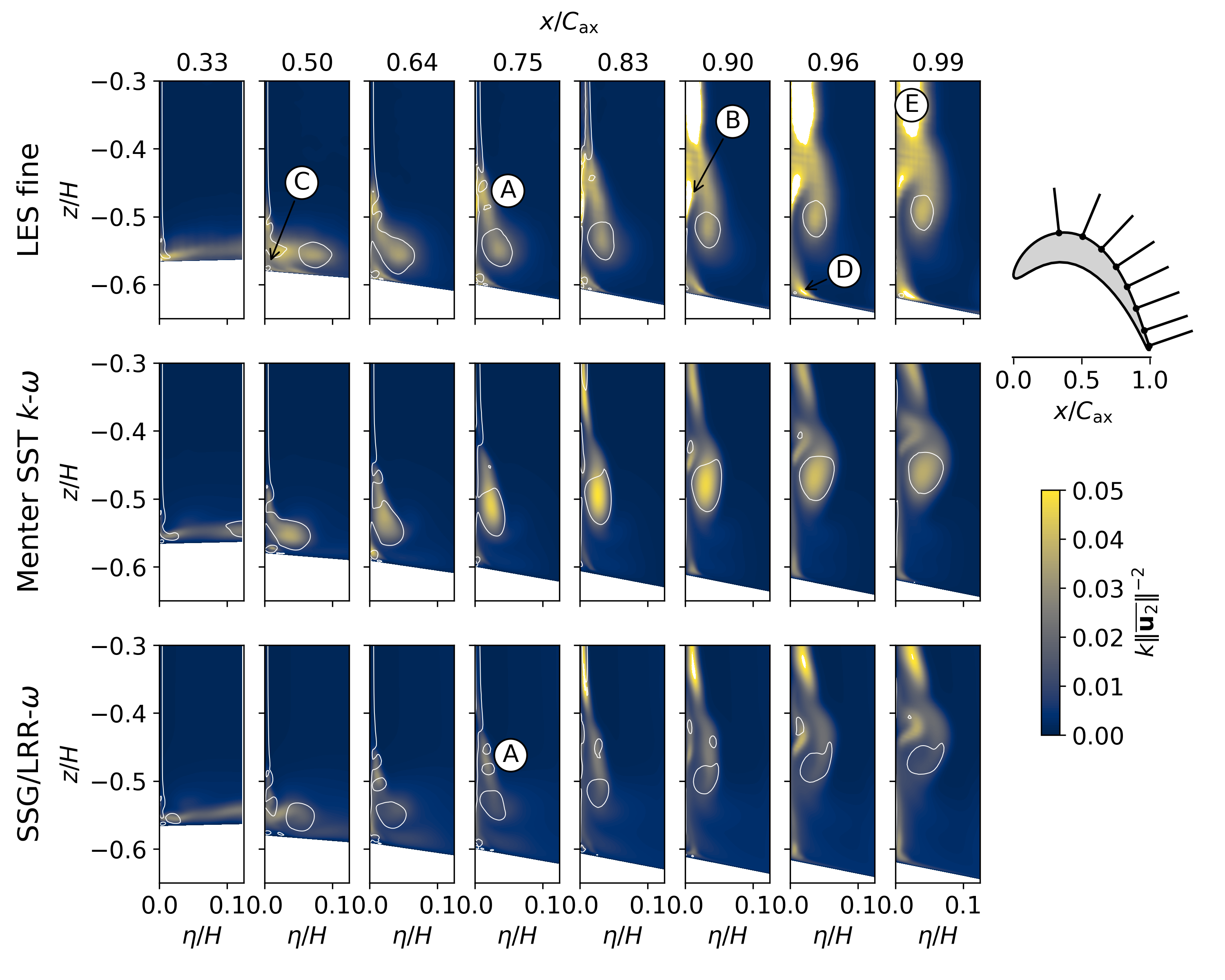}
    \caption{Turbulent kinetic energy $k$ in different cuts normal to blade surface for LES on fine mesh (\textit{top}), \MenterSST\ (\textit{middle}) and \SSGLRRw\ (\textit{bottom}); passage vortex core illustrated using $\overline{Q} C_{\mathrm{ax}}^2 / \Vert \overline{\mathbf{u}_2} \Vert^2 = 10$ contour line}
    \label{fig:boundaryLayerCuts_tke}
\end{figure}

Following the overview of the secondary flow system, we now turn to a more detailed analysis of the turbulence close to the suction side of the blade as potential root cause for the discrepancies in the wake losses discussed earlier.
For the midspan region, this has already been discussed extensively by Fard Afshar et al.~\cite{Afshar2022} and our results are consistent with their findings.
We, therefore, focus on the end wall region.
Fig.~\ref{fig:boundaryLayerCuts_tke} shows the turbulence kinetic energy $k$ in eight different wall-normal planes over the suction side surface whose positions are sketched in the schematic on the right.
Note, that the $z$-coordinate is normalised by the inflow channel height $H$ instead of the local channel height $h$ and that $\eta$ denotes the distance from the blade wall.
A single contour line of $\overline{Q} C_{\mathrm{ax}}^2 / \Vert \overline{\mathbf{u}_2} \Vert^2 = 10$ helps identify vortical structures in a more quantitative manner, with the most prominent being the \ac{PV}.
\ac{RANS} predicts the vortex to approach the blade wall more upstream, as can be seen in the cut at $\xbycax = 0.50$.
Subsequently, it lifts off the end wall more quickly leading to the discrepancy observed in the wake plane in Figs.~\ref{fig:wake_2d} and~\ref{fig:wake_2d_vorticity}.
A similar observation has been made by Marconcini et al.~\cite{Marconcini2019} for the T106A with parallel end walls.
The vortex core coincides with elevated $k$, which is qualitatively reproduced well by the \MenterSST\ model with a slight tendency to overpredict $k$ towards the trailing edge of the blade.
Curiously, the \SSGLRRw\ model fails to produce enough $k$ over the development of the \ac{PV} providing a possible explanation for the slower vortex decay in comparison with the \ac{LES} results.
Two other vortex cores can be observed above the \ac{PV} in the \ac{LES} at $\xbycax = 0.75$ (\circled{A}, also see Fig.~\ref{fig:endwall_flow_instantaneous} \circled{B}).
They are also produced by the \ac{DRSM} but again dissipate too slowly, as they can still be seen at $\xbycax = 0.96$.
In the \ac{LES} on the other hand, the vortices have disappeared by $\xbycax = 0.9$, where a patch of  high turbulence levels lifts off the blade surface (\circled{B}).
Moreover, a small corner vortex (\circled{C}) can be seen where the blade intersects the end wall.
It travels up the blade for a short distance before it is terminated again by the increasing turbulence intensity in this area (\circled{D}).
Qualitatively, this can also be found in the \ac{RANS}, again with significantly lower levels of $k$.
The largest discrepancy in terms of $k$ can be seen in the region of essentially two-dimensional flow (\circled{E}).
As discussed above, this is caused by the severe underprediction of the separation bubble by the \ac{RANS} models but has no significant influence on the discussion of the end wall flow.

\begin{figure}
    \centering
    \includegraphics[width=\textwidth]{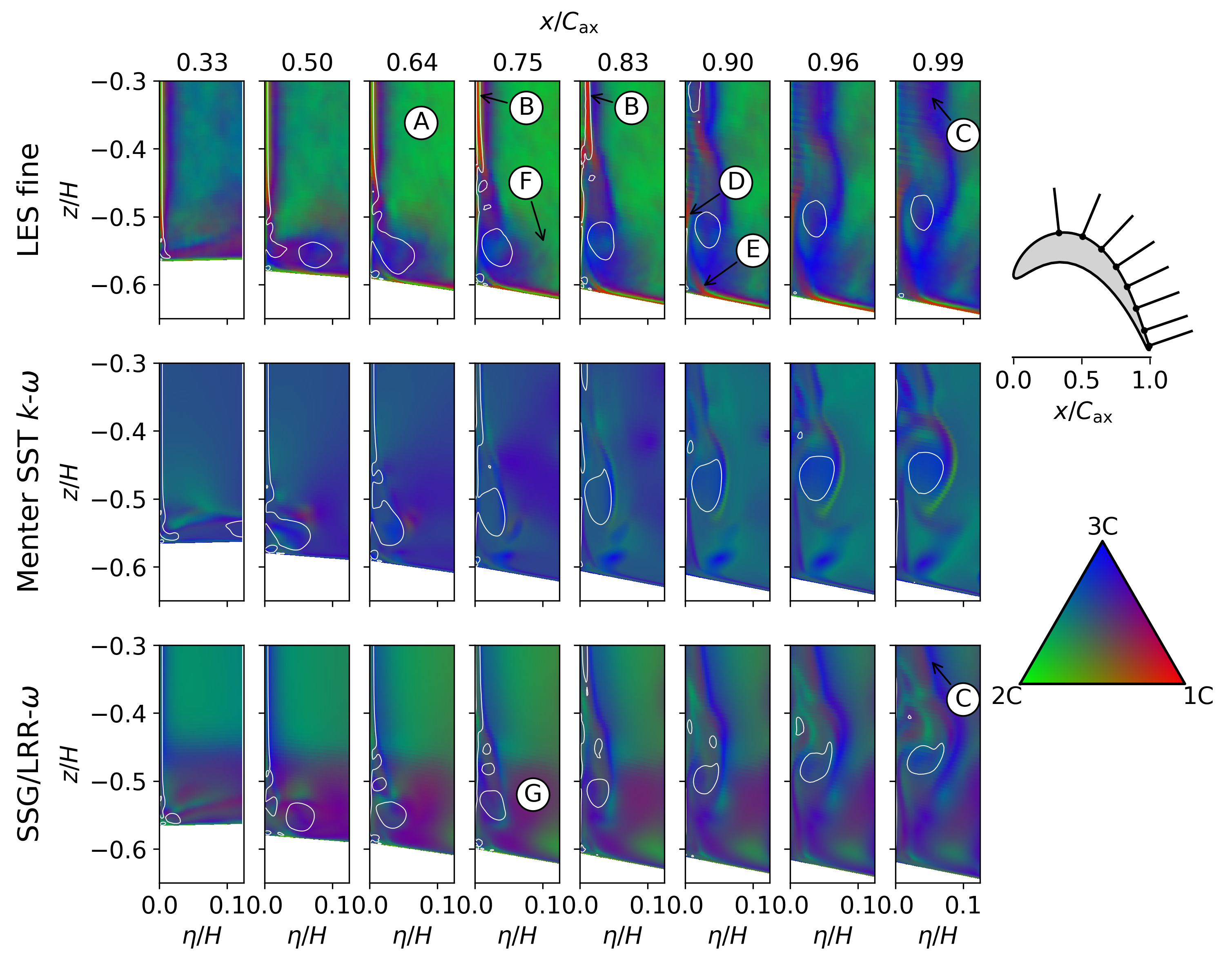}
    \caption{Reynolds stress tensor anisotropy visualised as RGB colours in different cuts normal to blade surface for LES on fine mesh (\textit{top}), \MenterSST\ (\textit{middle}) and \SSGLRRw\ (\textit{bottom}); passage vortex core illustrated using $\overline{Q} C_{\mathrm{ax}}^2 / \Vert \overline{\mathbf{u}_2} \Vert^2 = 10$ contour line}
    \label{fig:boundaryLayerCuts_anisotropy}
\end{figure}

We want to conclude our discussion of the secondary flow system with an analysis of the turbulence topology in terms of Reynolds stress anisotropy to add some facts to the common notion that \acp{DRSM} should outperform \acp{LEVM} in flow regions with significant turbulence anisotropy~\cite{Hanjalic2002}.
This discussion is facilitated by using the approach of Emory and Iaccarino~\cite{Emory2014} to spatially visualise the eigenvalues of the Reynolds stress anisotropy tensor by colour coding them as RGB channels.
Fig.~\ref{fig:boundaryLayerCuts_anisotropy} shows the same cut planes as above now coloured according to the turbulence anisotropy.
Each colour corresponds to a set of anisotropy tensor eigenvalues expressed as position in the barycentric triangle in the legend.
Starting our discussion with the freestream turbulence in the blade passage away from the end walls (\circled{A}), the colour map shows that the fluctuations transform from mixed two- and three-component turbulence towards the two-component (2C) corner (see also Fig.~\ref{fig:inflow_intensity_and_anisotropy}).
While this behaviour is reproduced reasonably well by the \ac{DRSM}, it cannot be expected from an \ac{LEVM} as the turbulence anisotropy equals the strain rate anisotropy through the linear Boussinesq approximation.
We nevertheless plot the results for the \MenterSST\ model for the sake of completeness.
Close to the blade, in the statistically two-dimensional part of the flow, the transitional separated shear layer can be observed in the $Q$-contour (\circled{B}) featuring turbulence distinctively close to the one-component limit, cf.~\cite{Afshar2022}.
After transition to turbulence, this area can be found significantly closer to isotropic (3C) turbulence (\circled{C}).
The \SSGLRRw\ model fails to reproduce this characteristic feature of the separation, and, as shown in Fig.~\ref{fig:midspan_streamlines}, the separation itself.
It only shows a reasonable match after transition (\circled{C}), yet, as discussed above, at much too low levels of turbulence.
Note, that the model is a high Reynolds number model with the simpler LRR pressure strain model~\cite{Launder1975} active close to solid walls.
While this choice contributes to the model's greater numerical stability compared to low Reynolds number approaches, it can be questioned in cases like the current.
A similar observation can be made for two more regions where near wall 1C-turbulence is lifted.
One is close to the end wall at the intersection with the blade wall (\circled{D}) and one within the upwards cross flow induced by the \ac{PV} on the blade surface (\circled{E}), both producing areas of increased turbulence (cf.~Fig.~\ref{fig:boundaryLayerCuts_tke} \circled{D} and \circled{B}).
While it was easy to associate the dominant \ac{PV} with increased $k$ in Fig.~\ref{fig:boundaryLayerCuts_tke}, the picture is not so clear in terms of turbulence anisotropy.
Generally, the turbulence is in a state close to isotropy in its area of influence, which is in agreement between \ac{LES} and \ac{DRSM}.
The \ac{LES}, however, shows this area to be confined relatively close to the blade wall and the region downstream of the \ac{PV} is quickly filled with 2C-turbulence from the freestream by its down-wash (\circled{F}).
Conversely, the \ac{DRSM} produces a large area of stresses in a rod-like state, there (\circled{G}).

In conclusion, none of the \ac{RANS} models is able to produce an overall satisfying picture of the turbulence in the secondary flow region.
But while the \MenterSST\ model is able to predict the levels of turbulence intensity within the \ac{PV} reasonably well, this is not the case with the  \SSGLRRw\ model and could be seen as a major shortcoming of this approach.
Neither model is able to accurately produce the high turbulence areas where fluid is lifted from the walls.
In terms of turbulence anisotropy, the \SSGLRRw\ cannot reproduce many of the distinctive features of the secondary flow system seen in the \ac{LES} results, despite being able to model the vortex stretching in the free stream.

\section{Conclusions}

We have presented a new well-resolved \ac{LES} dataset of the MTU T161 at $\Ma_{2,\mathrm{th}} = 0.6$, $\re_{2,\mathrm{th}} = 90000$ and $\alpha_1 = 41^\circ$ obtained with a high-order \ac{DG} method with a focus on the appropriate reproduction of inflow turbulent boundary layers and freestream turbulence.
Average blade loading and pressure loss distribution attributed to secondary flow features computed by the \ac{LES} agree well with the experiment and  underline the validity of the presented approach.
However, the case poses a challenge to \ac{RANS} models due to the laminar-to-turbulent transition at midspan and the complex secondary flow system at the end walls.
The analysis was conducted using an \ac{LEVM} and a \ac{DRSM}, both coupled with a state-of-the-art topology-independent transition model.
With the \ac{LES} reference at hand, boundary conditions could be reproduced consistently with the \ac{RANS} setup and  all differences between the simulation approaches can be attributed to physical modelling problems.
At midspan, no appropriate prediction of the separation-induced transition could be obtained leading to a too narrow wake of the blade.
Because the midspan flow solution is shown to only weakly effect on the secondary flow system, we could still obtain a meaningful assessment of the \ac{RANS} model performance in this region.
Both models overestimate local total pressure losses in the wake due to the \ac{PV} and \ac{TSV}, which can be traced back to the development of the vortex system beginning with the \ac{HSV} at the blade \ac{LE}.
Although the \ac{DRSM} shows a more appropriate representation of turbulence anisotropy throughout the flow field, it suffers from an underprediction of turbulence associated with large scale vortices, leading to too compact and persistent secondary flow structures.
In summary, we were able to identify important shortcomings in common \ac{RANS} approaches, which can be addressed in the future due to the improved consistency between \ac{LES} and \ac{RANS} setup in contrast to the traditional comparisons with experiments.

\section*{Declarations}

\begin{itemize}
\item Funding: No external funding was received for this work.
\item Conflict of interest: The authors declare that they have no conflict of interest.
\item Ethical approval: Not applicable.
\item Informed consent: Not applicable.
\item Authors' contributions: C.M. wrote the main manuscript text, prepared the figures, performed post-processing and the RANS simulations.
      M.B. developed the DG solver and wrote the part of the manuscript describing it.
      A.T. performed the LES and post-processing.
      B.K. contributed to the post-processing.
      M.F. contributed the experimental data.
      All authors reviewed the manuscript.
\end{itemize}

\bibliography{literature}


\end{document}